\iftrue
\documentclass[aps,prl,twocolumn,
			   groupedaddress,superscriptaddress,
			   amsfonts,amssymb,amsmath,
			   citeautoscript,
			   a4paper]{revtex4-2}
\else
\documentclass[aps,pra,preprint,
			   groupedaddress,superscriptaddress,
			   amsfonts,amssymb,amsmath,
			   citeautoscript,
			   a4paper]{revtex4-2}
\fi

\usepackage[pdftex]{hyperref}
\hypersetup{colorlinks,
			linkcolor={blue!75!black!80!yellow},
			citecolor={blue!75!black!80!yellow},
			urlcolor={blue!75!black!80!yellow},
			pdfstartview=FitH}
\usepackage{graphicx}
\usepackage{mathrsfs}
\usepackage{xspace}
\usepackage{braket}
\usepackage{xr}
\usepackage{gensymb} 
\usepackage{xcite}
\usepackage{xcolor,soul}
\usepackage{stmaryrd} 
\usepackage[UKenglish]{babel}
\usepackage{placeins} 
\usepackage{physics}

\usepackage{siunitx}
\DeclareSIUnit[number-unit-product=]\percent{\char`\%} 

\usepackage{txfonts}  
\usepackage{txfontsb} 

\makeatletter
\newcommand*{\addFileDependency}[1]{
  \typeout{(#1)}
  \@addtofilelist{#1}
  \IfFileExists{#1}{}{\typeout{No file #1.}}
}
\makeatother

\makeatletter
\renewcommand\@make@capt@title[2]{%
	\@ifx@empty\float@link{\@firstofone}{\expandafter\href\expandafter{\float@link}}%
	\sffamily{\textbf{#1}}\@caption@fignum@sep#2
}%

\makeatother

\renewcommand{\Re}{\operatorname{Re}}
\renewcommand{\Im}{\operatorname{Im}}
\newcommand{\iu}{\mathrm{i}}
\newcommand{\e}{\mathrm{e}}
\newcommand{\du}{\mathrm{d}}

\newcommand{\para}{\parallel\xspace}

\newcommand{\dpar}{d_{\parallel}}
\newcommand{\dperp}{d_{\perp}}
\newcommand{\jump}[1]{\llbracket #1 \rrbracket}

\newcommand{\nv}{\hat{\mathbf{n}}}

\newcommand{\epsz}{\varepsilon_0}
\newcommand{\eps}{\varepsilon}

\newcommand{\appropto}{\mathrel{\vcenter{
			\offinterlineskip\halign{\hfil$##$\cr
				\propto\cr\noalign{\kern2pt}\sim\cr\noalign{\kern-2pt}}}}}

\usepackage{textcomp} 
\usepackage{xifthen}
\usepackage{etoolbox}
\newboolean{togglecomments}
\newboolean{togglechanges}

\setboolean{togglecomments}{true}
\setboolean{togglechanges}{false}

\newcommand{\comment}[2]{%
    \ifbool{togglecomments}%
    {\textcolor{blue!70!black}{\small\textsf{%
    \textsuperscript{\textsc{\textsf{\MakeLowercase{#1}}}}%
    [#2]}}} 
    {}}     
\newcommand{\swap}[2]{\ifbool{togglechanges}
    {#2}  
    {\textcolor{red!70!black}{[#1]}\textrightarrow{}\textcolor{green!50!black}{[#2]}}}
\newcommand{\remove}[1]{\ifbool{togglechanges}
    {}    
    {\textcolor{red!70!black}{#1}}}
\newcommand{\inset}[1]{\ifbool{togglechanges}
    {#1}  
    {\textcolor{green!50!black}{#1}}}
\newcommand{\optional}[1]{\ifbool{togglechanges}
    {}    
    {\textcolor{yellow!50!orange!80!gray}{#1}}}

\newcommand{\citeremind}[1]{%
    [\textcolor{blue!75!black!80!yellow}{
        $\blacksquare$%
	    \ifthenelse{\isempty{#1}}
	        {}
	        {\textsuperscript{\tiny\textsf{#1}}}%
	}]\xspace}

\newcommand{\ie}{i.e.,\@\xspace}  

\newcommand{\eg}{e.g.,\@\xspace}

\newcommand{\hkuaffil}{\footnotesize Department of Physics and HK Institute of Quantum Science and Technology, The University of Hong Kong, Pokfulam, Hong Kong, China}
\newcommand{\denaffil}{\footnotesize Department of Electrical and Photonics Engineering, Technical University of Denmark, Kgs.\ Lyngby, Denmark}
\newcommand{\shaffil}{\footnotesize College of Optical-Electrical Information and Computer Engineering, University of Shanghai for Science and Technology, Shanghai 200093, China}
\newcommand{\genaffil}{\footnotesize Genuine Optronics Limited, Shanghai, China}

\begin{document}

\title{
Broadband measurement of Feibelman's quantum surface response functions
    }

\author{Zeling~Chen}
\thanks{Z.~C., S.~Y., and Z.~X. contributed equally to this work.}
\affiliation{\hkuaffil}
\author{Shu~Yang}
\thanks{Z.~C., S.~Y., and Z.~X. contributed equally to this work.}
\affiliation{\hkuaffil}
\author{Zetao~Xie}	
\thanks{Z.~C., S.~Y., and Z.~X. contributed equally to this work.}
\affiliation{\hkuaffil}
\author{Jinbing~Hu}
\affiliation{\hkuaffil}
\affiliation{\shaffil}
\author{Xudong~Zhang}
\affiliation{\hkuaffil}
\author{Yipu~Xia}
\affiliation{\hkuaffil}
\author{Yonggen~Shen}
\affiliation{\genaffil}
\author{Huirong~Su}
\affiliation{\genaffil}
\author{Maohai~Xie}
\affiliation{\hkuaffil}
\author{Thomas Christensen}
\affiliation{\denaffil}
\author{Yi~Yang}
\email{yiyg@hku.hk}
\affiliation{\hkuaffil}

\begin{abstract}
    The Feibelman $d$-parameter, a mesoscopic complement to the local bulk permittivity, describes quantum optical surface responses for interfaces, including nonlocality, spill-in and-out, and surface-enabled Landau damping. It has been incorporated into the macroscopic Maxwellian framework for convenient modeling and understanding of nanoscale electromagnetic phenomena, calling for the compilation of a $d$-parameter database for interfaces of interest in nano-optics. However, accurate first-principles calculations of $d$-parameters face computational challenges, whereas existing measurements of $d$-parameters are scarce and restricted to narrow spectral windows. We demonstrate a general broadband ellipsometric approach to measure $d$-parameters at a gold--air interface across the visible--ultraviolet regimes. Gold is found to spill-in and spill-out at different frequencies. We also observe gold's Bennett mode, a surface-dipole resonance associated with a pole of the $d$-parameter, around \SI{2.5}{\eV}.    
    Our measurements give rise to and are further validated by the passivity and Kramers--Kronig causality analysis of $d$-parameters.
    Our work advances the understanding of quantum surface response and may enable applications like enhanced electron field emission.
\end{abstract}

\maketitle

\begin{figure}[htbp]
	\centering
	\includegraphics[width=1\linewidth]{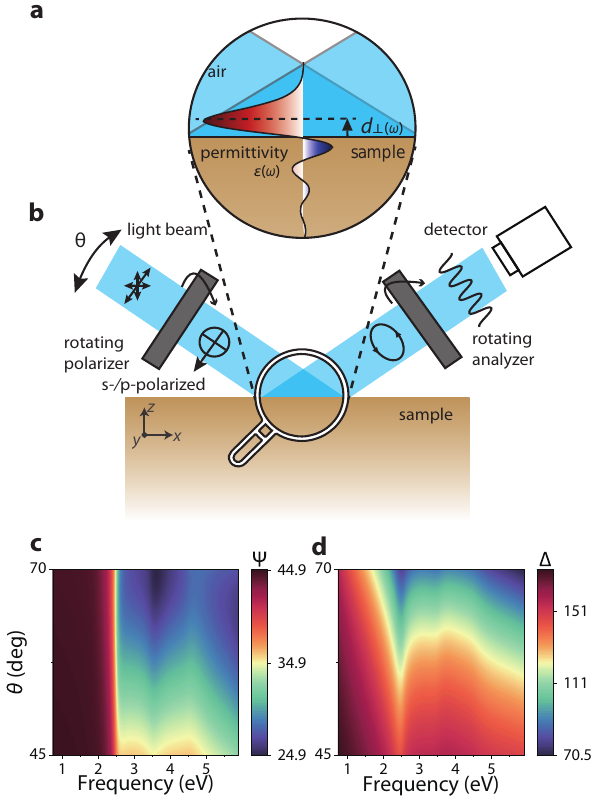}
	 \caption{%
            \textbf{Experimental setup and measured amplitude contrast and phase difference.}
            \textbf{a.}~Conceptual schematic of induced charge density at an interface. $\dperp$ is the centroid of the induced charge.
            \textbf{b.}~Ellipsometric setup.
            Incident light is linearly polarized under various incident angles $\theta$, reflected by the surface, and detected to obtain the amplitude contrast $\Psi(\omega,\theta)$ and phase difference $\Delta(\omega,\theta)$.
            \textbf{c--d.}~$\Psi(\omega,\theta)$ and $\Delta(\omega,\theta)$ of the gold--air interface averaged over all repeated measurements.
	    }
	\label{fig:framework}
\end{figure}

In classical electrodynamics, induced charges are assumed to be located at exactly the ionic interfaces. This assumption is valid at the macroscopic level but gradually fails in the microscopic limit. 
The deviation from the classical behavior is described by the Feibelman $d$-parameters (Fig.~\ref{fig:framework}a)~\cite{feibelman1982surface,feibelman1975microscopic,liebsch2013electronic}:
\begin{equation}
    \dperp(\omega)=\frac{\int_{-\infty}^\infty z\,\rho(z)\,\du z}{\int_{-\infty}^\infty \rho(z)\,\du z}, \qquad \dpar(\omega)=\frac{\int_{-\infty}^\infty z\,\partial_xJ(z)\, \du z}{\int_{-\infty}^\infty \partial_xJ(z)\, \du z}.
    \label{eq:d_definition}
\end{equation}
where $\dperp$ and $\dpar$ equal the centroids of the induced charge $\rho(z)$ and the normal derivative of the tangential current $J(z)$, respectively, at a planar interface
~\cite{mortensen2021mesoscopic,tserkezis2020applicability,menabde2022image,della2022orbital,yan2022efficient}.
The $d$-parameters are complex, dispersive (\ie frequency-dependent), and depend on the materials that make up the interface.
Conceptually, the $d$-parameters enable a comprehensive mesoscopic treatment of the quantum response effects at interfaces, including nonlocality~\cite{Boardman:1982a,Dominguez:2012,Raza:2015b,gonccalves2021quantum,boroviks2022extremely,ye2023nonlocal,aupiais2023ultrasmall}, electron spill-in/out~\cite{skjolstrup2018quantum,campos2019plasmonic,toscano2015resonance,khalid2020enhancing}, and surface-enabled Landau damping~\cite{yannouleas1992landau,molina2002oscillatory,yuan2008landau,khurgin2017landauACS} before the onset of quantum tunneling~\cite{zhu2014quantum,zhu2016quantum,deeb2023electrically} and size quantization~\cite{halperin1986quantum,monreal2013competition} at the atomic length scale.
Since inception, the study on the $d$-parameters has been carried out for about half a century~\cite{bagchi1977transverse,brodsky1980electrodynamics,apell1981simple,maniv1980electrodynamics,schaich1989nonlocal,del1981microscopic,langreth1989macroscopic,abeles1980ellipsometry,persson1984reference,liebsch1985violation,kempa1989nonlocal,tarriba1992collective}. 
During which the representative early experimental explorations had been the hunting of the multipole plasmon mode (also known as the Bennett mode~\cite{bennett1970influence}) in particular in simple Jellium-like metals~\cite{tsuei1990multipole,tsuei1991normal,sprunger1992normal,barman1998photoinduced,chiarello2000surface}.

Recently, the $d$-parameters can be included in the standard macroscopic Maxwell equations by a simple modification of the associated boundary conditions as self-consistent field discontinuities~\cite{yang2019general,yan2015projected}, enabling the analysis and modeling of nonclassical nanoscale ($\gtrsim\SI{1}{\nm}$) electromagnetic phenomena~\cite{kreibig1985optical, cottancin2006optical, grammatikopoulos2013self, berciaud2005observation, ciraci2012probing, scholl2012quantum, raza2013blueshift, alcaraz2018probing,yang2018numerical,baumberg2019extreme} with standard computational tools.
This utility and ease of implementation has been well-demonstrated by its adoption in state-of-the-art computational frameworks for nanophotonics, such as the boundary element method~\cite{hohenester2022nanoscale}, quasi-normal modes~\cite{tao2022quasinormal,zhou2022quasinormal}, transformational optics~\cite{yang2022transformation}, and fluctuational electrodynamics~\cite{zhang2024nanoscale}.
Building on these advances, recent work has firmly established that quantum surface effects, as described by the $d$-parameters, can strongly modify light-matter interactions across a wide variety of nanophotonic settings~\cite{christensen2017quantum,gonccalves2020plasmon,ciraci2019plasmonic,eriksen2024nonlocal,karanikolas2021quantum,babaze2022quantum,babaze2023dispersive}.
Given this, it is essential to develop theoretical and experimental methods for obtaining the $d$-parameters of technologically relevant optical interfaces, analogously to those used in determining the bulk permittivity.

On the theoretical side, time-dependent density functional theory within the jellium approximation has been widely applied to calculate $d$-parameters of simple alkali-like metals~\cite{feibelman1982surface,liebsch2013electronic,liebsch1987dynamical,liebsch1993surface,lang1970theory}.
However, the jellium treatment is inadequate for transition metals, and in particular, for the noble metals that are of key photonic and plasmonic importance, due to the non-negligible screening from lower-lying orbits~\cite{liebsch1993surface,feibelman1994comment,christensen2017quantum}. This, combined with the challenges in going beyond the jellium approximation in a well-controlled and computationally feasible way, calls for the development of experimental techniques to measure $d$-parameters directly.

Among the earliest experimental work toward this direction is the evidence of Bennett mode in silver obtained via the differential energy loss in low energy electron diffraction~\cite{moresco1996evidence}. 
Meanwhile, the observation of substantial quantum corrections in small nanospheres embedded in matrices enabled the extraction of $d$-parameters close to the nonretarded resonant frequency~\cite{charle1998surface}.
Analogous ideas were extended to the retarded regime via gap-plasmon nanoresonators, enabled by a quasi-normal-mode perturbation analysis~\cite{yang2019general}. 
Similarly, it has been shown that $d$-parameters can be extracted from the measured dispersion of highly-confined gap plasmons~\cite{boroviks2022extremely}.
More recently, DC-biased single nanoresonators led to the observation of spectral shift and linewidth modifications~\cite{zurak2024modulation}, implicitly demonstrating an electrical tunability of $d$-parameters.
The spectral ranges of these earlier measurements are all well below the nonretarded surface plasmon frequency of materials, and their narrowband nature requires modifying the geometry of structures to scan across frequencies.
As a result, the experimental tabulation of $d$-parameters remains highly challenging, especially in frequency ranges not covered by the resonant features of specific geometries.
This conundrum has thus hindered the observation of the widely-predicted Bennett mode~\cite{bennett1970influence,tsuei1990multipole} in a wide range of materials of plasmonic and photonic prominence and the confirmation of essential response function properties of $d$-parameters, such as passivity and the Kramers--Kronig (KK) relations~\cite{persson1983sum,persson1984reference,kempa1989nonlocal,liebsch1985violation}.

Here, we demonstrate a general ellipsometric approach to measuring $d$-parameters in a broadband manner using a gold--air interface in the visible and ultraviolet regimes spanning \SIrange{2}{6}{\eV}.
We confirm the feasibility of the approach using the Sobol$'$ sensitivity analysis and Monte Carlo simulations. 
Our measurements discover a surface-dipole Bennett mode near \SI{2.5}{\eV} and demonstrate that gold can spill either -in or -out, relative to the interface, depending on frequency. 
Finally, the broadband nature of our measurements enables the unprecedented direct experimental evidence and analysis of the KK relations and passivity of $d$-parameters.

Ellipsometry is a well-established technique for characterizing the bulk permittivity from a difference of reflection coefficients at a planar interface~\cite{fujiwara2007spectroscopic}. The experimental observables are $\Psi$ and $\Delta$ which give the amplitude contrast and phase difference between the $\mathrm{p}$- and $\mathrm{s}$-polarized reflection coefficients $r_\mathrm{p}$ and $r_\mathrm{s}$ via $\tan\Psi \e^{\iu\Delta}=r_\mathrm{p}/r_\mathrm{s}$. 
For a single planar interface comprising material 1 above and material 2 below, a set of quantum Fresnel equations (see Supp. Sec. S1) for $r_{\mathrm{p}}$ and $r_{\mathrm{s}}$ can be expressed through the Feibelman $d$-parameters, which incorporate the leading-order deviation from classical response~\cite{feibelman1982surface,kempa1985nonlocal,kempa1986calculation,gonccalves2020plasmon,apell1981simple,schaich1989nonlocal} (Fig.~\ref{fig:framework}b):%
\begin{equation}
\begin{aligned}
r_\mathrm{p}
&=\frac{\eps_\mathrm{2}k_{z,\mathrm{1}}-\eps_\mathrm{1}k_{z,\mathrm{2}}+(\eps_\mathrm{2}-\eps_\mathrm{1})(\mathrm{i}q^2\dperp-\mathrm{i}k_{z,\mathrm{1}}k_{z,\mathrm{2}}d_\parallel)}{\eps_\mathrm{2}k_{z,\mathrm{1}}+\eps_\mathrm{1}k_{z,\mathrm{2}}-(\eps_\mathrm{2}-\eps_\mathrm{1})(\mathrm{i}q^2\dperp+\mathrm{i}k_{z,\mathrm{1}}k_{z,\mathrm{2}}d_\parallel)},                      \\
r_\mathrm{s}
&=\frac{k_{z,\mathrm{1}}-k_{z,\mathrm{2}}+(\eps_\mathrm{2}-\eps_\mathrm{1})\mathrm{i}k_0^2d_\parallel}{k_{z,\mathrm{1}}+k_{z,\mathrm{2}}-(\eps_\mathrm{2}-\eps_\mathrm{1})\mathrm{i}k_0^2d_\parallel},
\label{eq:rsrp}
\end{aligned}
\end{equation}
where $k_0 \equiv \omega/c$, $k_j \equiv \sqrt{\eps_j}k_\mathrm{0}$ for $ j \in \{\mathrm{1},\mathrm{2}\}$, $k_{z,j}\equiv \sqrt{k_j^2-q^2}$, and $q$ is the in-plane wavevector. Evidently, the classical Fresnel equations are restored when $\dperp=\dpar=0$.

Figure~\ref{fig:framework}c and d show the measured $\Psi$ and $\Delta$ in our experiment on a gold--air interface (Supp. Sec. S6). The template-stripped gold samples are of uniform Au$\mathrm{(111)}$ orientation (Supp. Sec. S5.A) with a surface roughness (rms) of \SI{0.311}{\nm} (Supp. Sec. S5.B).
The gold layer is effectively bulk, considering its \SI{100}{\nm} thickness that substantially exceeds the skin depth in the measured frequency region.
For frequencies $\hbar\omega\gtrapprox\SI{2}{\eV}$, both $\Psi$ and $\Delta$ exhibit rich variations, which suggest the possibility of using the quantum Fresnel equations [Eq.~\eqref{eq:rsrp}] to measure $d$-parameters.
In fact, this possibility was suggested speculatively in early discussions~\cite{schaich1988anomalous,kempa1985nonlocal,kempa1985importance,abeles1980ellipsometry}, but was seen as impractically challenging at fixed incident angles due to the difficulty of accurately quantifying the relatively minor quantum correction contributions without resonant enhancement.

\begin{figure*}[htbp]
	\centering
	\includegraphics[width=1\linewidth]{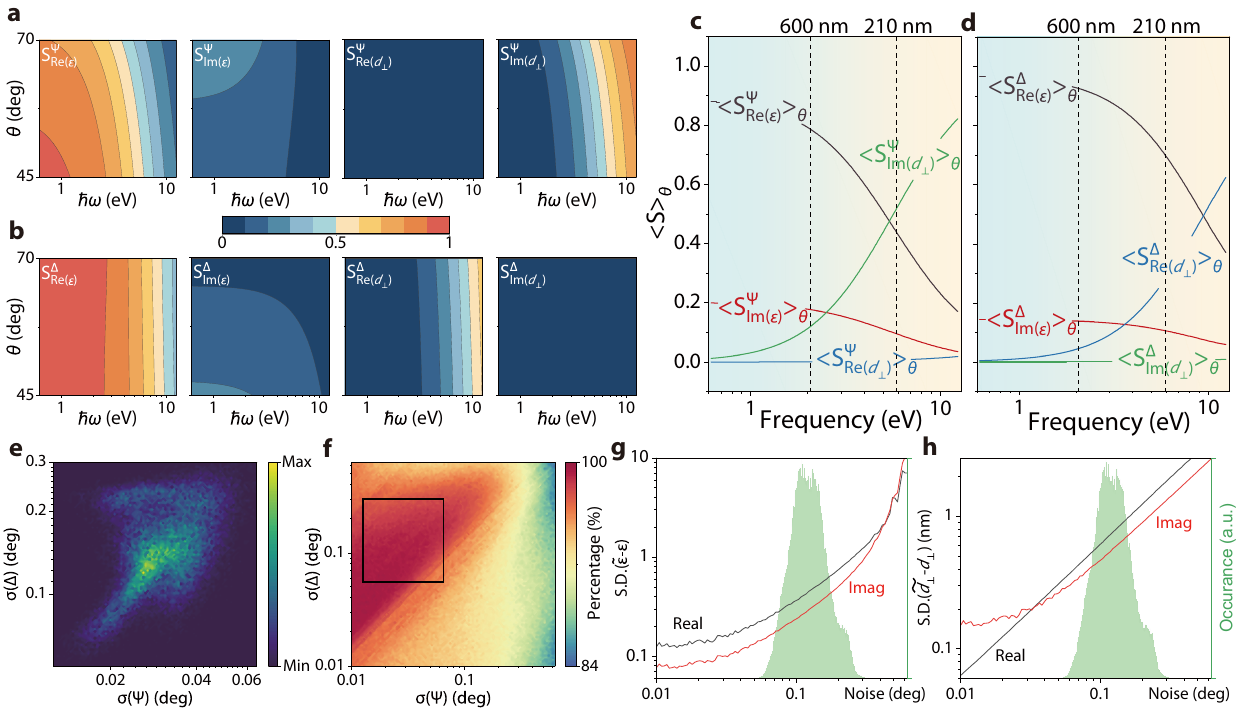}
	 \caption{%
	 	\textbf{Measurement feasibility confirmed by Sobol$'$ sensitivity analysis and Monte Carlo simulations.}
	 	\textbf{a--b.}~Total-effect indices $S^{\Psi,\Delta}_{\eps,\dperp}$ describe the decomposed contributions from the bulk permittivity $\eps(\omega)$ and surface $\dperp(\omega)$ to observables $\Psi(\theta,\omega)$ (a) and $\Delta(\theta,\omega)$ (b) at various incident angles.
            \textbf{c--d.}~Angle-integrated total-effect indices $\langle S^{\Psi,\Delta}_{\eps,\dperp}\rangle_\theta$. 
            The gradient shading from blue to yellow indicates the increased and decreased contribution of $\dperp(\omega)$ and $\eps(\omega)$, respectively, towards higher frequencies.
            The vertical dashed lines show the data analysis window, limited by the sensitivity of the $d$-parameter (low-frequency cutoff) and laser tunability (high-frequency cutoff).
            \textbf{e.}~Experimental noise in ellipsometry observables $\Psi$ and $\Delta$.
            \textbf{f.}~Percentage of Monte Carlo simulations where the ground truth of permittivity and $d$-parameter falls within the $3\sigma$ confidence interval ($\sigma$ is the standard deviation) under various noise levels.
            The experimental noise statistics (e) lie within the black box in f.
            \textbf{g--h.} Standard deviations between fitting and ground-truth of $\eps$ (f) and $\dperp$ (g) of a single sampling.
            The histogram (shaded green in g--h) shows the experimental noise levels at various frequencies and incident angles.
	    }
	\label{fig:sobol_mc}
\end{figure*}

\begin{figure*}[htbp]
	\centering
	\includegraphics[width=1\linewidth]{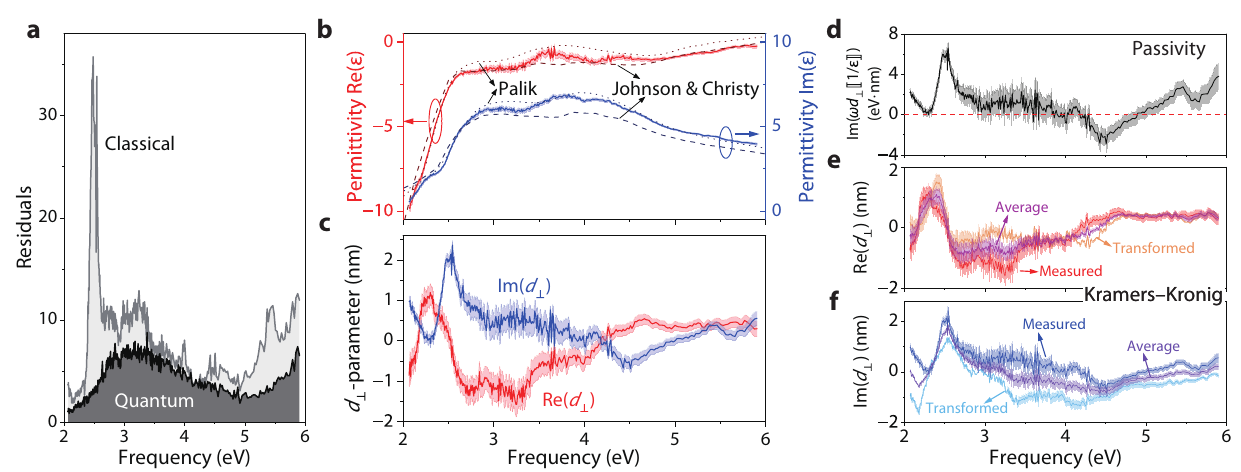}
	 \caption{%
	 	\textbf{Measured $\dperp$ of the gold--air interface and its passivity and Kramers--Kronig analysis.}
            \textbf{a.}~The quantum Fresnel equations Eq.~\eqref{eq:rsrp} reduces fitting residuals especially around $\SI{2.5}{\eV}$, suggesting a resonant behavior of $\dperp(\omega)$.
	 	\textbf{b--c.}~Measured bulk $\eps(\omega)$ and surface $\dperp(\omega)$. 
            Real and imaginary parts are in red and blue, respectively, and shadings denote $3\sigma$ confidence interval derived from fitting.
            In b, dotted lines and dashed lines denote the~\citet{palik1998handbook} and~\citet{johnson1972optical} data, respectively.
            In c, sharp variation of $\dperp(\omega)$ appears near \SI{2.5}{\eV} in accordance with the residual reduction in a.
            \textbf{d.}~The measured data satisfy the passivity requirement [Eq.~\eqref{eq:passivity_dperp}] at most frequencies.
            \textbf{e--f.}~The broadband measurement enables self-consistent Kramers--Kronig (KK) transformation between real (e) and imaginary (f) parts of $\dperp(\omega)$, which includes the measured data, the KK-transformed data, and their average.
	 	}
	\label{fig:data}
\end{figure*}

Using the gold--air interface as an example, we confirm the feasibility of ellipsometric determination of the $d$-parameters, using two independent methods, \ie the Sobol$'$ sensitivity analysis and a Monte Carlo approach.
In these methods (Fig.~\ref{fig:sobol_mc}) and the data analysis below (Fig.~\ref{fig:data}), we assume $d_\para=0$, which is generally expected at simple, charge-neutral interfaces~\cite{liebsch2013electronic,apell1981simple}.

First, we apply the variance-based Sobol$'$ sensitivity analysis~\cite{sobol2001global,saltelli2010variance,Herman2017}
to compare the relative weights of $\dperp$ and $\eps$ to observables and identify the frequency regimes where the ellipsometric method can apply.
Intuitively, the Sobol$'$ analysis quantitatively attributes output variance to input variance (see Supp. Sec. S4.A for details). Of central importance is the total-effect index measuring the influence of an input, taking into account the interactions of that variable with all other inputs.
Figure~\ref{fig:sobol_mc}a--b shows the total-effect indices $S_\mu^\nu$, where $\mu=\left\{\Re \eps, \Im \eps, \Re \dperp, \Im\dperp\right\}$ and $\nu=\left\{\Psi,\Delta\right\}$, to characterize the total influence of the bulk permittivity and $d_\perp$-parameter under a range of incident angles and frequencies.
Corresponding angle-averaged indices are shown in Fig.~\ref{fig:sobol_mc}c and d.
In the Drude regime ($\hbar\omega\lessapprox\SI{2}{\eV}$), the contributions from the bulk permittivity predominate the response of $\Psi$ and $\Delta$, complicating the extraction of $d$-parameters via ellipsometry. 
At higher frequencies ($\hbar\omega\gtrapprox\SI{2}{\eV}$), however, the influence of $\eps$ and $\dperp$ become comparable, demonstrating the feasibility of ellipsometric $d$-parameter evaluation in this regime.
A complementary sensitivity analysis (Supp. Sec. S4.B) confirms this conclusion.
Notably, this regime covers a range above the screened plasma frequency where dielectric screening is non-negligible and where both theoretical and experimental results for $d$-parameters have thus-far been lacking.

Second, we apply a Monte Carlo approach to confirm that the quantum Fresnel equations allow robust extraction of $\dperp$ also in the presence of experimental noise in $\Psi$ and $\Delta$ (Fig~\ref{fig:sobol_mc}e and green histograms in Fig.~\ref{fig:sobol_mc}g--h).
In the Monte Carlo simulation, we randomly choose the ground-truth combinations of $\eps$ and $\dperp$, add noises of different levels to observables $\Psi$ and $\Delta$, and perform nonlinear least squares fitting to obtain estimates $\Tilde{\eps}$ and $\Tilde{\dperp}$ across a Monte Carlo ensemble (Supp. Sec. S7.B).
Within a $10^4$-ensemble dataset for each point in Fig.~\ref{fig:sobol_mc}f and $10^6$-ensemble dataset in Fig.~\ref{fig:sobol_mc}g--h, the ground truth falls within $3\sigma$ of the estimates with $99\%$ likelihood (Fig.~\ref{fig:sobol_mc}f).
At an experimental $(\Psi,\Delta)$-noise level of $\SI{0.1}{\degree}$ (see the histogram in Fig.~\ref{fig:sobol_mc}g--h), the single-sample standard deviations in $\eps$ is near 0.3 (Fig.~\ref{fig:sobol_mc}g), while that of $\dperp$ is around \SI{5}{\angstrom} (Fig.~\ref{fig:sobol_mc}h), both of which can be further reduced by repeated measurements.

Figure~\ref{fig:data} shows the measurement data and associated analysis. 
We first compare the fitting residuals (Supp. Sec. S7.A) of the quantum Fresnel equations [Eq.~\eqref{eq:rsrp}] with those of their classical counterpart in Fig.~\ref{fig:data}a.
A substantial residual reduction, especially pronounced near $\SI{2.5}{\eV}$, is achieved with the quantum Fresnel equations, enabling us to simultaneously extract the bulk permittivity of gold (Fig.~\ref{fig:data}b) and the $\dperp$ of the gold--air interface (Fig.~\ref{fig:data}c).
The thereby inferred bulk permittivity agrees well with the widely-used tabulated data from~\citet{palik1998handbook} and from~\citet{johnson1972optical} (dotted and dashed lines in Fig.~\ref{fig:data}a).
Moreover, the confidence intervals from the experimental fitting (shading in Fig.~\ref{fig:data}b--c) are also corroborated by post-measurement Monte Carlo analysis (Supp. Sec. S7.C).

We emphasize two key features observed in the extracted $\dperp(\omega)$ spectrum.
First, in addition to the observation of charge spill-in ($\Re\dperp<0$) at low frequencies ($\lessapprox\SI{2}{\eV}$), in agreement with numerous earlier observations of nonclassical blueshifts in gold nanostructure spectra~\cite{ciraci2012probing,cottancin2006optical,yang2019general}, we observe spill-out ($\Re\dperp>0$) in the high-frequency tail.
The onset of this spill-out regime is around \SI{2.5}{\eV}, slightly below the screened plasma frequency $\omega_\mathrm{p}^{*}\approx\SI{2.8}{\eV}$.
This observation of a high-frequency spill-out region in complement to a low-frequency spill-in region, is, in fact, a natural consequence and a first implicit experimental evidence of a KK-related sum rule [Eq.\eqref{eq:KKrelation}], stating that $\int_0^\infty \mathrm{d}{\omega}\,\Re\dperp(\omega) = 0$.
Meanwhile, $\Im \dperp$ is non-negative, except for a narrow and likely spurious region near \SI{4.5}{eV}.

Second, we discover a surface-dipole mode in gold, also known as the Bennett mode~\cite{bennett1970influence,tsuei1990multipole,baghramyan2021laplacian,tsuei1991normal,yan2015hydrodynamic}.
The mode is associated with the clear resonant peak feature\inset{~\cite{feibelman1982surface,liebsch1987dynamical,kempa1988nonlocal}} in $\Im\dperp(\omega)$ and an associated zero-crossing of $\Re\dperp(\omega)$ appears near \SIrange{2.4}{2.5}{\eV}, \ie with a pole of $\dperp(\omega)$ at complex $\omega$.
The quality factor of the Bennett mode is measured as 13 from the width of $\Im{\dperp}$.
Notably, this narrow spectral region is accompanied by a substantial residual reduction (Fig.~\ref{fig:data}a).
The appearance of Bennett mode below the screened plasma frequency has been widely predicted  theoretically~\cite{christensen2017quantum,liebsch2013electronic,echarri2021optical,liebsch1995influence} and probed experimentally~\cite{tsuei1990multipole,tsuei1991normal,barman1998photoinduced,chiarello2000surface,moresco1996evidence} in simple metals and silver, but its experimental confirmation has been lacking for gold thus far.
We obtain consistent results in repeated time-separated measurements (Supp. Sec. S8.A) and across multiple samples (Supp. Sec. S8.B).

The measured $d$-parameters are several times larger than previous calculations, which could reflect imperfections in earlier theoretical models and/or in our sample and measurement conditions; nevertheless, the magnitude of $\Re \dperp$ in our Au data is similar to that of Ag~\cite{charle1998surface} near but below their screened plasma frequencies, respectively.

Additionally, by relaxing the assumption of $\dpar=0$, we performed another type of fitting on the quantity $\dperp-\dpar$ (See Refs.~\cite{kempa1985nonlocal,schaich1989nonlocal,feibelman1982surface,liebsch2013electronic} and Fig.S10 and Supp. Sec. S9) because $\mathrm{(111)}$ noble metals may host Shockley surface states that contribute to a nonzero $\dpar$~\cite{feibelman1982surface,echarri2021optical}. 
The extracted $\dperp-\dpar$ is in quantitative agreement with $\dperp(\omega)$ in Fig.~\ref{fig:data}c (which assumes $\dpar(\omega)=0$), indicating the consistency and robustness of the data. 
The fine angular resolution in incident angles (Fig.~\ref{fig:framework}c and d) generates a sufficiently large number of observations, which play a critical role in extracting the $d$-parameters with unprecedented accuracy in past inaccessible frequency regimes. 

To carefully check the consistency of our extracted $\dperp(\omega)$ values, we perform two additional validations: based, respectively, on requirements of passivity and causality.
To this end, we first derive the constraints imposed by passivity by initially re-formulating the $d$-parameters via a nonclassical surface susceptibility (Supp. Sec. S3)
\begin{align}
    \boldsymbol{\chi}(\mathbf{r}) 
    =
    \left[ \dperp\jump{1/\eps}\hat{\mathbf{n}}\hat{\mathbf{n}}^{\mathrm{T}}-\dpar\jump{\eps} \big(\mathbf{\hat{I}}-\hat{\mathbf{n}}\hat{\mathbf{n}}^{\mathrm{T}}\big)\right]\delta_{\partial\Omega}(\mathbf{r}),
    \label{eq:susceptibility}
\end{align}
which enables us to directly establish a passivity constraint by requiring that the influence of the $d$-parameters is strictly dissipative at all frequencies, \ie that $\omega\boldsymbol{\chi}$ be positive-definite (Supp. Sec. S3)%
\begin{subequations}%
\begin{align}
    &\Im(\omega\dperp\jump{1/\eps})\geq 0 ,\label{eq:passivity_dperp}
    \\
    & \Im(\omega\dpar\jump{\eps})\leq 0 .\label{eq:passivity_dpara}
\end{align}%
\label{eq:passivity}%
\end{subequations}%
Here, $\jump{\mathbf{f}}\equiv \mathbf{f}^+ -  \mathbf{f}^-$ denotes the discontinuity of a scalar or vectorial field $\mathbf{f}$ across an interface $\partial\Omega$ with outward normal $\nv$ and $\mathbf{f}^\pm \equiv (\textbf{r}_{\partial\Omega}\pm 0^ + \hat{\mathbf{n}})$.
$\delta_{\partial\Omega}(\mathbf{r})$ denotes a surface delta function associated with the interface $\partial\Omega$. 
The perpendicular and parallel component of $\mathbf{f}$ is defined by $f_\perp\equiv \hat{\mathbf{n}}\cdot \mathbf{f}$ and $\mathbf{f}_\parallel \equiv (\mathbf{\hat{I}}-\mathbf{\hat{n}}\mathbf{\hat{n}}^{\mathrm{T}})\mathbf{f}$, $\mathbf{\hat{I}}$ is a unit-matrix and $\textbf{r} \in \mathbb{R}^3$.
In Eq.~\eqref{eq:passivity}, the bulk permittivities across the interface are assumed isotropic.

We evaluate the passivity requirement using the ellipsometrically extracted $\dperp(\omega)$ values in Fig.~\ref{fig:data}d.
Except near the spurious region where $\Im\dperp(\omega)<0$, around \SI{4.5}{\eV}, the passivity constraint is satisfied, also in accordance with our quantum surface energy dissipation calculation (See Supp. Sec. S11).
The violation at \SI{4.5}{\eV} could \eg be a result of the neglect of contributions from $\dpar$ (see Fig.S10 and Supp. Sec. S9) or, considering the violation's modest magnitude and extent, as due to experimental imperfections. (See summary in Supp. Sec. S12)

Next, since the $d$-parameters must be causal functions, they are subject to a set of Kramers--Kronig (KK) relations (Supp. Sec. S2)~\cite{persson1983sum,persson1984reference}%
 \begin{subequations}
\begin{align}
    &\Re d_{\perp}( \omega) =\frac{2}{\pi}\int_{0}^{\infty}{\mathrm{d}\omega'\,\frac{\omega'\Im d_{\perp}( \omega') }{\omega'^{2}-\omega^2}},\label{eq:KKRe}
    \\
    &\Im d_{\perp}( \omega) =-\frac{2}{\pi}\int_{0}^{\infty}{\mathrm{d}\omega'\,\frac{\omega\Re d_{\perp}( \omega') }{\omega'^{2}-\omega^2}}.\label{eq:KKIm}
\end{align}%
\label{eq:KKrelation}%
\end{subequations}%
This KK relation was first obtained by~\citet{persson1983sum} decades ago but has remained experimentally elusive.
The KK relations offer a path to validate the broadband consistency of extracted $\dperp(\omega)$, by comparing the ``bare'' fit values to the corresponding values obtained by a subsequent KK transformation (with suitable low- and high-frequency extrapolation; see Supp. Sec. S10).%
To this end, we compare the directly extracted values of $\dperp(\omega)$ to their KK-transformation, as well as their mutual average, in Fig.~\ref{fig:data}e--f.
The key features of the spectra are consistent across these variants: in particular, the key resonant feature of the $d$-parameters at \SIrange{2.4}{2.5}{\eV} are accordant.
%

\begin{figure}[htbp]
	\centering
	\includegraphics[width=\linewidth]{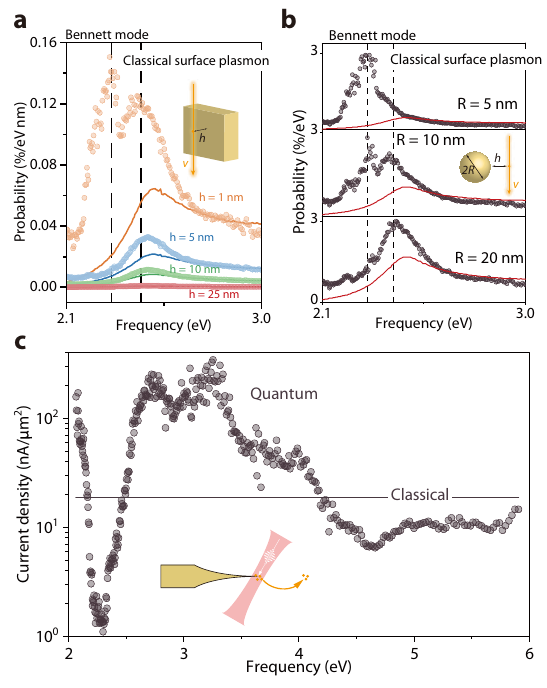}
	  \caption{%
	  \textbf{Quantum corrections and Bennett mode in electron energy loss and field emission.}
        \textbf{a.}~Enhancement of energy loss of a \SI{10}{\keV} electron moving parallel near a planar gold--air half-space at various separations. 
        \textbf{b.}~Enhancement of energy loss of a \SI{10}{\keV} electron moving near a gold sphere of different radii at separations of \SI{1}{\nm}.
        \textbf{c.}~Enhancement and suppression of field emission under quantum correction. The peak field amplitude assumes $F_0 = \SI{15}{V/nm}$, a repetition rate $f_\mathrm{R}=\SI{100}{\MHz}$, and a full-width at half-maximum pulse duration of $\SI{10}{\fs}$.
        Solid lines and dots represent the classical and quantum-corrected results, respectively.
        The measured permittivity and $\dperp$ are both incorporated in the three calculations.
	    }
	\label{fig:cases}
\end{figure}

Finally, we examine the implications of our measurement, particularly the Bennett mode, across three complementary scenarios, namely for the electron energy loss of a half-space (Fig.~\ref{fig:cases}a) and a sphere (Fig.~\ref{fig:cases}b), and field emission (Fig.~\ref{fig:cases}c). 
First, we examine the energy loss of a single electron above a gold half-space in Fig.~\ref{fig:cases}a (Supp. Sec. S13.C)~\cite{gonccalves2023interrogating}. 
The energy loss increases as the separation reduces in both classical and quantum calculations, whereas the latter exhibits stronger enhancement.
More strikingly, at the small separation of $h=\SI{1}{nm}$, a double-peak lineshape emerges where the high-frequency peak results from the excitation of surface plasmons as inherited from the classical treatment, whereas the new low-frequency peak is a direct consequence of the Bennett mode.

Second, we discuss the energy loss of a single electron near a gold sphere in Fig.~\ref{fig:cases}b (Supp. Sec. S13.D)~\cite{gonccalves2023interrogating}. 
Surprisingly, with the sphere raduis gradually deceasing, the dominnant plasmonic modes shift from the classical surface plasmonic mode ($R=\SI{20}{\nm}$) to the Bennett mode ($R=\SI{5}{\nm}$). At the moderate radius $R=\SI{10}{\nm}$, two plasmonic modes share the equal weight, showing a double-peak lineshape in the spectrum. 

Third, we investigate the quantum surface effects on field emission (Supp. Sec. S14), where the consequence of $d$-parameters has not been explored previously.
The nonzero $\dperp$ contributes an out-of-plane surface dipole density of $\boldsymbol{\pi}\equiv \epsz\dperp\jump{E_\perp}\hat{\mathbf{n}}$, which in turn shifts the work function by $\Delta \phi = -e\pi/\epsz$~\cite{leung2003relationship,ibach2006physics} due to surface charge redistribution induced by the field discontinuity in response to an ultrafast pulse (Fig.~\ref{fig:cases}c inset).
By inserting the field-modified work function in the Fowler--Nordheim equation, we obtain the changing of current density due to the surface response shown in Fig~\ref{fig:cases}c. 
The result predicts that it is possible to enhance or suppress the field emission by optically pumping a field emitter at a frequency where electrons spill-in or spill-out, respectively.

In conclusion, we demonstrate an ellipsometric approach for broadband measurement of the Feibelman $d$-parameters. 
The approach is verified and complemented by passivity and causality consistency tests.
Gold is found to \mbox{spill-in} and \mbox{-out} in different frequency regimes and to exhibit a surface-dipole Bennett mode around \SI{2.5}{\eV}.
Nevertheless, our measurement still has its limitations due to imperfections in surface adsorption, potential surface charging during sample peel-off, and surface roughness (see Supp. Sec. S12).
These findings and limitations jointly call for future refined measurements at various interfaces, such as those consisting of metals with distinctly non-Drude dispersion, doped seminconductors~\cite{west2010searching}, and even liquids~\cite{lv2024photomolecular}, where the method demonstrated here can be generally applied.
The methodology developed here can be adopted directly in commercial (and potential under vacuum conditions) ellipsometers and could inspire the development of tailored ellipsometric modulation and demodulation schemes~\cite{fujiwara2007spectroscopic} for community-wide access to $d$-parameter measurements and lead to their associated comprehensive cataloging.
Such efforts would be useful in furthering the understanding of mesoscopic quantum effects in condensed matter physics, surface chemistry, vacuum electronics, and nanophotonics.




\emph{Acknowledgments.}
We thank Yuan~Cao, Frankie~Y.~F.~Chan, Sangyeon~Cho, Yichen~Feng, F.~Javier~Garc\'{i}a~de~Abajo, \mbox{Francisco~J.~Garc\'{i}a-Vidal}, James~Hilfiker, Zemeng~Lin, Tran~Trung~Luu, N.~Asger~Mortensen, Owen~D.~Miller, Tom~Tiwald, and Ruichuan~Zhang for experimental help and stimulating discussions.


    
\bibliographystyle{apsrev4-2}
\bibliography{main}

\end{document}


\title{SUPPLEMENTRAY INFORMATION\\
Broadband measurement of Feibelman's quantum surface response functions
    }

\author{Zeling~Chen}
\thanks{Z.~C., S.~Y., and Z.~X. contributed equally to this work.}
\affiliation{\hkuaffil}
\author{Shu~Yang}
\thanks{Z.~C., S.~Y., and Z.~X. contributed equally to this work.}

\affiliation{\hkuaffil}
\author{Zetao~Xie}	
\thanks{Z.~C., S.~Y., and Z.~X. contributed equally to this work.}
\affiliation{\hkuaffil}
\author{Jinbing~Hu}
\affiliation{\hkuaffil}
\affiliation{\shaffil}
\author{Xudong~Zhang}
\affiliation{\hkuaffil}
\author{Yipu~Xia}
\affiliation{\hkuaffil}
\author{Yonggen~Shen}
\affiliation{\genaffil}
\author{Huirong~Su}
\affiliation{\genaffil}
\author{Maohai~Xie}
\affiliation{\hkuaffil}
\author{Thomas Christensen}
\affiliation{\denaffil}
\author{Yi~Yang}
\email{yiyg@hku.hk}
\affiliation{\hkuaffil}

\maketitle

\noindent{\textbf{\textsf{CONTENTS}}}\\ 
\twocolumngrid
\begingroup 
    \let\bfseries\relax 
    \deactivateaddvspace 
    \deactivatetocsubsections 
    \tableofcontents
\endgroup
\onecolumngrid

\section{The quantum Fresnel equations}
\label{smsec:fresnel}
Here we derive Fresnel equations with quantum surface corrections.
We first rewrite the mesoscopic boundary condition where $d$-parameters mediate self-consistent field discontinuities~\cite{yang2019general}. 
\begin{subequations}\label{smeq:mesoscopic_bcs}
\begin{align}
    &\jump{D_\perp} 
    = -\iu\omega^{-1}\nablav_{\para}\cdot\mathbf{K}
    = \dpar\nablav_\para\cdot\jump{\mathbf{D}_\para},
    \label{smeq:mesoscopic_bcs_Dperp}
    \\
    &\jump{B_\perp} 
    = 0,
    \label{smeq:mesoscopic_bcs_Bperp}
    \\
    &\jump{\mathbf{E}_\para} 
    = -\epsz^{-1}\nablav_\para\pi = -\dperp\nablav_\para\jump{E_\perp},
    \label{smeq:mesoscopic_bcs_Epara}
    \\
    &\jump{\mathbf{H}_\para} 
    = \mathbf{K}\times \nv 
    = \iu\omega\dpar\jump{\mathbf{D}_\para}\times\nv.
    \label{smeq:mesoscopic_bcs_Hpara}
\end{align}%
\end{subequations}%
Here, $\jump{\mathbf{f}}\equiv \mathbf{f}^+ -  \mathbf{f}^-$ denotes the discontinuity of a field $\mathbf{f}$ (an electric $\textbf{E}$, displacement $\textbf{D}$, magnetic $\textbf{B}$ or magnetizing $\textbf{H}$ field with determined frequency $\omega$) across an interface $\partial\Omega$ with outward normal $\nv$. $\delta_{\partial\Omega}(\mathbf{r})$ denotes a surface delta function associated with the interface $\partial\Omega$. The field $\mathbf{f}^+$ and $\mathbf{f}^-$ are defined as $\mathbf{f}^\pm \equiv \left(\textbf{r}_{\partial\Omega}\pm 0^ + \hat{\mathbf{n}}\right)$. The perpendicular and parallel component of the fields $\mathbf{f}$ is defined by $f_\perp\equiv \hat{\mathbf{n}}\cdot \mathbf{f}$ and $\mathbf{f}_\parallel \equiv \left(\mathbf{\hat{I}}-\mathbf{\hat{n}}\mathbf{\hat{n}}^{\mathrm{T}}\right)\mathbf{f}$, $\mathbf{\hat{I}}$ is a unit-matrix and $\textbf{r} \in \mathbb{R}^3$.
With these boundary conditions, we rederive the reflection and transmission coefficients of a planar interface. We adopt the same coordinate system as in Main Text Figure 1, \ie material 1 at the top and material 2 at the bottom.

\subsection{p-polarized waves}\label{smsec:fresnel_p}

First, we focus on the transverse magnetic solutions:
\begin{subequations}\label{smeq:field_1}
\begin{align}
    &\textbf{H}_\mathrm{1}=
        \left(\e^{-\iu k_{z,\mathrm{1}}z}+r^{(H)}_\mathrm{p}\e^{\iu k_{z,\mathrm{1}}z}\right)\e^{\iu \left(qx-\omega t\right)}\,\hat{\textbf{y}},
    \\
    &\textbf{E}_\mathrm{1}=\left[E_{x,\mathrm{1}}(z)\hat{\textbf{x}}+E_{z,\mathrm{1}}(z)\hat{\textbf{z}}\right]\e^{\iu \left(qx-\omega t\right)},
\end{align}%
\end{subequations}%
in the half-space of the material 1 ($z>0$), and
\begin{subequations}\label{smeq:field_2}
\begin{align}
    &\textbf{H}_\mathrm{2}=
        t^{(H)}_\mathrm{p}\e^{-\iu k_{z,\mathrm{2}}z}\e^{\iu \left(qx-\omega t\right)}\hat{\textbf{y}},
    \\
    &\textbf{E}_\mathrm{2}=\left[E_{x,\mathrm{2}}(z)\hat{\textbf{x}}+E_{z,\mathrm{2}}(z)\hat{\textbf{z}}\right]\e^{\iu \left(qx-\omega t\right)},
\end{align}%
\end{subequations}%
in the half-space of the material 2 ($z<0$). $q$ is the in-plane wavevector and $k_j \equiv \sqrt{\eps_j}k_\mathrm{0}$ for $ j \in \{\mathrm{1},\mathrm{2}\}$ are the out-of-plane wavevectors. Here, the superscript ``$(H)$'' stands for magnetic fields.
Using Maxwell's equations, the amplitudes of the electric fields can be obtained as
\begin{subequations}\label{smeq:field_ampli}
\begin{align}
    &E_{x,\mathrm{1}}(z)=\frac{-k_{z,\mathrm{1}}}{\omega \epsz \eps_\mathrm{1}}\left(\e^{-\iu k_{z,\mathrm{1}}z}-r^{(H)}_\mathrm{p}\e^{\iu k_{z,\mathrm{1}}z}\right),
    \\
    &E_{z,\mathrm{1}}(z)=\frac{-q}{\omega \epsz \eps_\mathrm{1}}\left(\e^{-\iu k_{z,\mathrm{1}}z}+r^{(H)}_\mathrm{p}\e^{\iu k_{z,\mathrm{1}}z}\right),
    \\
    &E_{x,\mathrm{2}}(z)=\frac{-t^{(H)}_\mathrm{p}k_{z,\mathrm{2}}}{\omega \epsz \eps_\mathrm{2}}\e^{-\iu k_{z,\mathrm{2}}z},
    \\
    &E_{z,\mathrm{2}}(z)=\frac{-t^{(H)}_\mathrm{p}q}{\omega \epsz \eps_\mathrm{2}}\e^{-\iu k_{z,\mathrm{2}}z}.
\end{align}%
\end{subequations}%

At the planar interface, substituting Eqs.~\eqref{smeq:field_ampli} into the boundary conditions in Eqs.~\eqref{smeq:mesoscopic_bcs}, we get
\begin{subequations}
\begin{align}
    &\frac{k_{z,\mathrm{1}}}{\eps_\mathrm{1}}\left(r^{(H)}_\mathrm{p}-1\right)+\frac{k_{z,\mathrm{2}}}{\eps_\mathrm{2}}t^{(H)}_\mathrm{p}=\iu q^2\dperp\left[\frac{1}{\eps_\mathrm{1}}\left(1+r^{(H)}_\mathrm{p}\right)-\frac{1}{\eps_\mathrm{2}}t^{(H)}_\mathrm{p}\right],
    \\
    &1+r^{(H)}_\mathrm{p}-t^{(H)}_\mathrm{p}=-\iu \dpar \left[k_{z,\mathrm{1}}\left(r^{(H)}_\mathrm{p}-1\right)+k_{z,\mathrm{2}}t^{(H)}_\mathrm{p}\right].
\end{align}%
\end{subequations}%

With some algebraic derivations, one can obtain the quantum reflection and transmission coefficient for $\mathrm{p}$-polarized wave, and translate in terms of electric fields with $r_\mathrm{p}\equiv r^{(H)}_\mathrm{p}$ and $t_\mathrm{p} \equiv \sqrt{\frac{\eps_\mathrm{1}}{\eps_\mathrm{2}}}t^{(H)}_\mathrm{p}$:
\begin{subequations}\label{smeq:coefficient_p_full}
\begin{align}
    &r_\mathrm{p}=\frac{\eps_\mathrm{2}k_{z,\mathrm{1}}-\eps_\mathrm{1}k_{z,\mathrm{2}}+\left(\eps_\mathrm{2}-\eps_\mathrm{1}\right)\left(\iu q^2\dperp-\iu k_{z,\mathrm{1}}k_{z,\mathrm{2}}\dpar\right)+\left(\eps_\mathrm{2}k_{z,\mathrm{2}}-\eps_\mathrm{1}k_{z,\mathrm{1}}\right)q^2\dperp\dpar}{\eps_\mathrm{2}k_{z,\mathrm{1}}+\eps_\mathrm{1}k_{z,\mathrm{2}}-\left(\eps_\mathrm{2}-\eps_\mathrm{1}\right)\left(\iu q^2\dperp+\iu k_{z,\mathrm{1}}k_{z,\mathrm{2}}\dpar\right)-\left(\eps_\mathrm{2}k_{z,\mathrm{2}}+\eps_\mathrm{1}k_{z,\mathrm{1}}\right)q^2\dperp\dpar},
    \label{smeq:rp_full}
    \\
    &t_\mathrm{p}=\sqrt{\frac{\eps_\mathrm{1}}{\eps_\mathrm{2}}}\frac{1}{1-\iu k_{z,\mathrm{2}}\dpar }\left[\left(1+\iu k_{z,\mathrm{1}}\dpar\right)r_\mathrm{p}-\iu k_{z,\mathrm{1}}\dpar+1\right].
\end{align}%
\end{subequations}%
If we only keep the leading linear order of $kd_{\perp,\parallel}$, the expressions reduce to
\begin{subequations}\label{smeq:coefficient_p_first}
\begin{align}
    &r_\mathrm{p}=\frac{\eps_\mathrm{2}k_{z,\mathrm{1}}-\eps_\mathrm{1}k_{z,\mathrm{2}}+\left(\eps_\mathrm{2}-\eps_\mathrm{1}\right)\left(\iu q^2\dperp-\iu k_{z,\mathrm{1}}k_{z,\mathrm{2}}\dpar\right)}{\eps_\mathrm{2}k_{z,\mathrm{1}}+\eps_\mathrm{1}k_{z,\mathrm{2}}-\left(\eps_\mathrm{2}-\eps_\mathrm{1}\right)\left(\iu q^2\dperp+\iu k_{z,\mathrm{1}}k_{z,\mathrm{2}}\dpar\right)},
    \label{smeq:rp}
    \\
    &t_\mathrm{p}=\frac{2\sqrt{\eps_\mathrm{1}\eps_\mathrm{2}}k_{z,\mathrm{1}}}{\eps_\mathrm{2}k_{z,\mathrm{1}}+\eps_\mathrm{1}k_{z,\mathrm{2}}-\left(\eps_\mathrm{2}-\eps_\mathrm{1}\right)\left(\iu q^2\dperp+\iu k_{z,\mathrm{1}}k_{z,\mathrm{2}}\dpar\right)}.
\end{align}%
\end{subequations}%

\subsection{s-polarized waves}
\label{smsec:fresnel_s}

The derivation of quantum Fresnel reflection and transmission coefficients of$\mathrm{s}$-polarized waves is similar to that of the $\mathrm{p}$-polarized wave. The coefficients can be defined for the transverse electric fields. By applying the nonclassical boundary conditions, we get:
\begin{subequations}\label{smeq:coefficient_s}
\begin{align}
    &r_\mathrm{s}=\frac{k_{z,\mathrm{1}}-k_{z,\mathrm{2}}+\left(\eps_\mathrm{2}-\eps_\mathrm{1}\right)\mathrm{i}k_0^2d_\parallel}{k_{z,\mathrm{1}}+k_{z,\mathrm{2}}-\left(\eps_\mathrm{2}-\eps_\mathrm{1}\right)\mathrm{i}k_0^2d_\parallel},\label{smeq:rs}
    \\
    &t_\mathrm{s}=\frac{2k_{z,\mathrm{1}}}{k_{z,\mathrm{1}}+k_{z,\mathrm{2}}-\left(\eps_\mathrm{2}-\eps_\mathrm{1}\right)\mathrm{i}k_0^2d_\parallel}.
\end{align}%
\end{subequations}%

\section{Kramers--Kronig relation and sum rules}
\label{smsec:KKrelation}

As response functions, $\dperp$ naturally satisfy a set of Kramers--Kronig (KK) relations~\cite{persson1983sum}:
 \begin{subequations}
\begin{align}
    &\Re\left[\dperp\left(\omega\right)\right]=\frac{2}{\pi}\int_{0}^{\infty}{\mathrm{d}\omega'\,\frac{\omega'\Im\left[\dperp\left( \omega'\right)\right]}{\omega'^{2}-\omega^2}},\label{smeq:KKR}
    \\
    &\Im\left[\dperp\left(\omega\right)\right]=-\frac{2}{\pi}\int_{0}^{\infty}{\mathrm{d}\omega'\,\frac{\omega\Re\left[\dperp\left( \omega'\right)\right]}{\omega'^{2}-\omega^2}}.\label{smeq:KKI}
\end{align}%
\label{smeq:KK}
\end{subequations}%

Obtaining sum rules for $\dperp$ then follows directly from the KK relations~\cite{persson1983sum}.
For $\Re(\dperp)$, multiplying $\omega$ to Eq.~\eqref{smeq:KKI} and taking the limit of $\omega\to \infty$, we have
\begin{subequations}\label{smeq:sumrules_deriv}
    \begin{align}
        \int^{\infty}_0 \diff{\omega}\, \Re\left[ \dperp\left(\omega\right)\right]=\frac{\pi}{2}\lim_{\omega\to \infty}\omega\Im{\dperp\left(\omega\right)}. \label{smeq:sumrules_deriv_re}
    \end{align}
\end{subequations}
Using the asymptotic property that $\dperp\left(\omega\right) \sim \omega^{-2}$ when $\omega\to \infty$~\cite{persson1983sum}, we can simplify the sum rule of $\Re\dperp$ in Eq.~\eqref{smeq:sumrules_deriv_re} to
\begin{align}
    &\int^{\infty}_0 \diff{\omega}\, \Re\left[ \dperp\left(\omega\right)\right]=0.
    \label{smeq:sumrules_re} 
\end{align}
For $\Im(\dperp)$, applying Eq.~\eqref{smeq:KKR} at zero frequency produces
\begin{align}
    \int^{\infty}_0 \diff{\omega}\, \frac{\Im\left[\dperp\left(\omega\right)\right]}{\omega} = \frac{\pi}{2}\dperp(0).\label{smeq:sumrule_Im}
\end{align}

\section{Passivity constraints for $d$-parameters}
\label{smsec:passivity}

The nonclassical surface dipole and current density can be recast through the introduction of a nonclassical surface susceptibility, thereby enabling the imposition of passivity constraints on the $d$-parameters.
From Eqs.~\eqref{smeq:mesoscopic_bcs_Epara} and ~\eqref{smeq:mesoscopic_bcs_Dperp}, the surface dipole $\boldsymbol{\pi}$ and current $\mathbf{K}$ can be expressed as~\cite{yang2019general}
\begin{subequations}\label{smeq:dipole_surrent_define}
\begin{align}
    &\boldsymbol{\pi}(\mathbf{r}_{\partial\Omega})\equiv \epsz\dperp\hat{\mathbf{n}}\hat{\mathbf{n}}^\mathrm{T}\jump{\mathbf{E}}, \label{smeq:dipole_define}\\
    &\mathbf{K}(\mathbf{r}_{\partial\Omega})\equiv \iu\omega\dpar\left(\mathbf{\hat{I}}-\hat{\mathbf{n}}\hat{\mathbf{n}}^\mathrm{T}\right)\jump{\mathbf{D}}.
\end{align}
\end{subequations}
The induced current is equivalent to a polarization $\iu\omega^{-1}\mathbf{K}$, allowing us to combine the two contributions into a single ``nonclassical'' polarization term
\begin{equation}\label{smeq:ncpolarization}
    \mathbf{P}^{\mathrm{nc}}(\mathbf{r}) = \left[\boldsymbol{\pi}(\mathbf{r})+\iu\omega^{-1}\mathbf{K}(\mathbf{r})\right]\delta_{\partial\Omega}(\mathbf{r}).
\end{equation}

Next, we assume a thin vacuum layer sandwiched between two materials with thickness $t$ and volume $\partial\Omega^t$ to precisely characterize the interface $\partial\Omega$, that is $\partial\Omega^t\in \left\{ \mathbf{r}\in\mathbb{R}^3 \mid \exists \mathbf{r}_{\partial\Omega}\in \partial\Omega~\mathrm{st}. \left|\big(\mathbf{r}-\mathbf{r}_{\partial\Omega}\big)\cdot \mathbf{\hat{n}}\right| \leq t/2 \right\}$. Associated with an indicator function $\theta_{\partial\Omega^t}(\mathbf{r})=1$ for $\mathbf{r}\in \mathbf{r}_{\partial\Omega^t}$ and $\theta_{\partial\Omega^t}(\mathbf{r})=0$ elsewhere, the layer is related to the interface represented by a surface delta function,
\begin{equation}\label{smeq:surface_delta}
    \delta_{\mathbf{r}_{\partial\Omega}}(\mathbf{r})\equiv \lim_{t\to 0+}t^{-1}\theta_{\partial\Omega^t}(\mathbf{r}).
\end{equation}
Thus the nonclassical polarization inside the layer is
\begin{equation}\label{smeq:polar_noncla}
    \mathbf{P}\big(\mathbf{r}\in \partial\Omega^t\big)=\left[\boldsymbol{\pi}+\iu\omega^{-1}\mathbf{K}(\mathbf{r}) \right]/t.
\end{equation}
For the points outside this region, the polarization is simply the classical one, $\mathbf{P}\big(\mathbf{r}\in\mathbb{R}^3 \backslash \partial\Omega^t\big)=\epsz\chi^{\mathrm{c}}(\mathbf{r})\mathbf{E}(\mathbf{r})$, where $\chi^{\mathrm{c}}$ is the classical susceptibility. 

Now due to the redefinition of $\partial\Omega^t$, the notions of the field in/outside have changed to $f^\pm=f\big(\mathbf{r}_{\partial\Omega}\pm \frac{1}{2} t^{+} \hat{\mathbf{n}}\big)$. Notice that the $d$-parameters framework is a leading-order picture. Thus, the process that swapping $\mathbf{E}$ for $\mathbf{E}^\mathrm{c}$ and $\mathbf{D}$ for $\mathbf{D}^\mathrm{c}$ is second order which is negligible since the classical parts $\mathbf{E}^\mathrm{c}$ and $\mathbf{D}^\mathrm{c}$ are only the first order of the quantum-corrected fields $\mathbf{E}$ and $\mathbf{D}$. This will refer to
\begin{subequations}\label{smeq:polar_curr_class}
\begin{align}
    &\boldsymbol{\pi}(\mathbf{r}_{\partial\Omega})= \epsz\dperp\hat{\mathbf{n}}\hat{\mathbf{n}}^\mathrm{T}\jump{\mathbf{E}^{\mathrm{c}}}, \label{smeq:polar_class}\\
    &\mathbf{K}(\mathbf{r}_{\partial\Omega})= \iu\omega\dpar\left(\mathbf{\hat{I}}-\hat{\mathbf{n}}\hat{\mathbf{n}}^\mathrm{T}\right)\jump{\mathbf{D}^{\mathrm{c}}}.\label{smeq:curr_class}
\end{align}
\end{subequations}
We can substitute the fields $\mathbf{E}^{\mathrm{c}\pm}$ to the fields at the interface $\partial \Omega$  classically which can be bridged by $\partial\Omega^t$. First, the normal $\mathbf{D}^\mathrm{c}$ field is continuous, \ie $D^{\mathrm{c}-}_\perp=D^{\mathrm{c}}_\perp=D^{\mathrm{c}+}_\perp$. Thus, we have $E^{\mathrm{c}\pm}_\perp=E^{\mathrm{c}}_\perp/\eps^{\mathrm{c}\pm}$ with $\eps^{\mathrm{c}\pm} \equiv \eps^\mathrm{c}\big(\mathbf{r}_{\partial\Omega}\pm \frac{1}{2} t^{+} \hat{\mathbf{n}}\big)$. We can substitute $\hat{\mathbf{n}}\hat{\mathbf{n}}^\mathrm{T}\jump{\mathbf{E}^{\mathrm{c}}}$ for $\hat{\mathbf{n}}\hat{\mathbf{n}}^\mathrm{T}\jump{1/\eps^{\mathrm{c}}}\mathbf{E}$ in Eq.~\eqref{smeq:polar_class}. Next, the tangential fields $\mathbf{E}^{\mathrm{c}}$ is continuous, \ie $\mathbf{E}^{\mathrm{c}-}_\parallel=\mathbf{E}^{\mathrm{c}}_\parallel=\mathbf{E}^{\mathrm{c}+}_\parallel$. 
Recalling the relation $\mathbf{D}^{\mathrm{c}\pm}_\parallel=\epsz\eps^{\mathrm{c}\pm}\mathbf{E}^\mathrm{c}_\parallel$, we substitute $\left(\mathbf{\hat{I}}-\hat{\mathbf{n}}\hat{\mathbf{n}}^\mathrm{T}\right)\jump{\mathbf{D}^{\mathrm{c}}}$ for $\left(\mathbf{\hat{I}}-\hat{\mathbf{n}}\hat{\mathbf{n}}^\mathrm{T}\right)\epsz\jump{\eps^{\mathrm{c}}}\mathbf{E}^\mathrm{c} $ in Eq.~\eqref{smeq:curr_class}. 
Inserting these two relations back in Eq.~\eqref{smeq:polar_noncla}, and substituting $\mathbf{E}^\mathrm{c}$ for the quantum-corrected field $\mathbf{E}$ (neglecting again the second-order error), we find
\begin{equation}\label{smeq:polar_substi}
    \mathbf{P}\big(\mathbf{r}\in \partial\Omega^t\big)=\epsz t^{-1}\left[ \dperp\jump{1/\eps^\mathrm{c}}\hat{\mathbf{n}}\hat{\mathbf{n}}^{\mathrm{T}}-\dpar\jump{\eps^\mathrm{c}}\left(\mathbf{\hat{I}}-\hat{\mathbf{n}}\hat{\mathbf{n}}^{\mathrm{T}}\right)\right]\mathbf{E}(\mathbf{r}).
\end{equation}
Now we can directly identify the nonclassical susceptibility $\boldsymbol{\chi}^{\mathrm{nc}}$ as  
\begin{align}
    \boldsymbol{\chi}^\mathrm{nc}(\mathbf{r}) = t^{-1}\left[ \dperp\jump{1/\eps^\mathrm{c}}\hat{\mathbf{n}}\hat{\mathbf{n}}^{\mathrm{T}}-\dpar\jump{\eps^\mathrm{c}}\left(\mathbf{\hat{I}}-\hat{\mathbf{n}}\hat{\mathbf{n}}^{\mathrm{T}}\right)\right].
    \label{smeq:susceptibility_t}
\end{align}
In the $t\rightarrow 0^+$ limit, we can then finally restore the surface delta function to obtain
\begin{align}
    \boldsymbol{\chi}^\mathrm{nc}(\mathbf{r}) = \left[ \dperp\jump{1/\eps^\mathrm{c}}\hat{\mathbf{n}}\hat{\mathbf{n}}^{\mathrm{T}}-\dpar\jump{\eps^\mathrm{c}}\left(\mathbf{\hat{I}}-\hat{\mathbf{n}}\hat{\mathbf{n}}^{\mathrm{T}}\right)\right]\delta_{\partial\Omega}(\mathbf{r}).
    \label{smeq:susceptibility}
\end{align}

Note that the nonclassical susceptibility is anisotropic and remains a local quantity. 
The classical and nonclassical susceptibilities will not be simultaneously nonzero, since the classical susceptibility $\chi^\mathrm{c}(\mathbf{r})$ is nonzero in $\mathbb{R}^3\backslash\partial \Omega^t$ and $\boldsymbol{\chi}^\mathrm{nc}$ is nonzero only in $\partial \Omega^t$.
Similar to the bulk permittivity, we can determine the constraints of passivity. We focus on the rate of energy absorption (dissipated power) in the nonclassical layer $\partial\Omega^t$: $\partial_tu=\frac{1}{2}\Re\int_{\partial\Omega^t}\diff^3\mathbf{r}\,\mathbf{E}^\dagger \mathbf{J}$. Recalling that $\mathbf{J}= -\iu\omega\mathbf{P}=-\iu\omega\epsz\boldsymbol{\chi}^{\mathrm{nc}}\mathbf{E}$, we get
\begin{equation}\label{smeq:passi_energy}
    \partial_tu=\frac{\epsz}{2}\int_{\partial\Omega^t}\diff^3\mathbf{\rv}\,\Im\left[\mathbf{E^\dagger}(\mathbf{r})\left(\omega\boldsymbol{\chi}^{\mathrm{nc}}\right)\mathbf{E}(\mathbf{r})\right],
\end{equation}
where there is the possibility that $\omega$ is a complex quantity with $\Im\omega>0$ to ensure causality. This must be zero or positive for any electric field $\mathbf{E(r)}$ at any position inside the layer $\mathbf{r}\in\partial\Omega^t$ to ensure a locally dissipative environment. Accordingly, the imaginary part of the operator $\omega\boldsymbol{\chi}^{\mathrm{nc}}$ must be positive-semidefinite, \ie $\Im\mathbf{f}^\dagger\left(\omega\boldsymbol{\chi}^{\mathrm{nc}}\right)\mathbf{f}\geq 0$ for any $\mathbf{f}\in\mathbb{C}^3$. Since $\boldsymbol{\chi}^\mathrm{nc}$ is diagonal, we can finally cast the constraints from passivity as
\begin{subequations}
\begin{align}
    &\Im\left(\omega\dperp\jump{1/\eps^\mathrm{c}}\right)\geq 0 ,\label{smeq:passivity_dperp}
    \\
    & \Im\left(\omega\dpar\jump{\eps^\mathrm{c}}\right)\leq 0 .\label{smeq:passivity_dpara}
\end{align}%
\label{smeq:passivity}
\end{subequations}%

\section{Model sensitivity analysis}\label{smsec:model_sensi}

We theoretically verify the contribution of $\dperp$ in the quantum Fresnel law to demonstrate that it can be extracted in principle. 
Considering an arbitrary model, the relative contribution (or weight) of various input parameters can be studied by calculating how sensitive the outputs are to the inputs. The sensitivity indices can be in several forms such as first-order (outputs to a single input), second-order (outputs to two inputs with potential correlation), and total-order (outputs to the total effect of a single input, including the first-order effect, higher-order effect of a single input). 

In this section, we provide two methods to identify and determine the valid frequency regimes where the measurement of $\dperp$ can be successful.
The first is the comprehensive Sobol$'$ variance-based sensitivity analysis and the second is a simple method using first-order partial differentiation. The former can calculate all kinds of sensitivity indices, whereas the latter only considers the first-order effects. 

\subsection{The Sobol$'$ sensitivity analysis}
\label{smsec:sobol}
The Sobol$'$ sensitivity analysis is based on the decomposition of the input parameters variance to the variance of output observables, which not only evaluate the sensitivity of individual parameters but also account for the interactions of multiple parameters~\cite{sobol2001global,saltelli2002making,saltelli2010variance,Iwanaga2022,Herman2017}.
It uses the Monte Carlo method to sample various input parameter combinations. As long as the sample size is large enough (to be ergodic), all potential parameter combinations can be achieved. Based on the combination of all inputs and corresponding outputs, we can statistically calculate their variance, respectively.

The input arguments $\eps$ and $\dperp$ are represented as $\mu$, the outputs $\nu$ which are the observables $\Psi$ and $\Delta$, and they are related through the quantum Fresnel equations as $\nu = f(\mu)$. The variances of output variables can be decomposed into the variance of inputs,
\begin{equation}
    \Var (\nu) = \sum_{i}V_{\mu_i}+\sum_{i<j}V_{\mu_i \mu_j}+\cdots+V_{\mu_1\mu_2\mu_3\mu_4},
    \label{smeq:var_decompose}
\end{equation}
where $V_{\mu_i}=\Var_{\mu_i}\left(E_{\mu_{\sim i}}\left(\nu\,\vline\,\mu_i\right)\right)$, $V_{\mu_i\mu_j}=\Var_{\mu_i\mu_j}\left(E_{\mu_{\sim ij}}\left(\nu\,\vline\,\mu_i,\mu_j\right)\right)$, and so on. $\Var$ denotes the variance, $E$ is the expectation value, and $\mu_{\sim i}$ denotes the inputs arguments without $\mu_i$. In such a way, one can define the first-order index as
\begin{equation}
    s_{\mu_i}^\nu=\frac{V_{\mu_i}}{\Var(\nu)},
    \label{smeq:sobol_first_order}
\end{equation}
which indicates the individual influence of an input $\mu_i$ on the output $\nu$. 
The total-effect index, which describes the contribution of an input with its interactions with the output variance taken into account, can be given as
\begin{equation}
    S_{\mu_i}^\nu=\frac{E_{\mu_{\sim i}}\left(\Var_{\mu_i}(\nu \,\vline\, \mu_{\sim i})\right)}{\mathrm{Var}(\nu)}.
    \label{smeq:sobol_total_order}
\end{equation}

\begin{figure*}[htbp]
    \centering
    \includegraphics[width=1\linewidth]{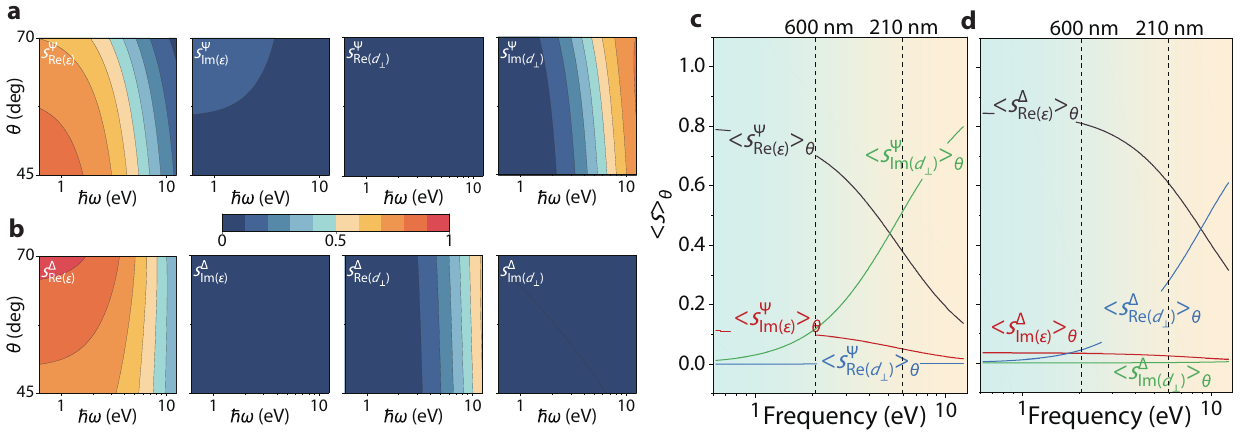}
    \caption{\textbf{First-order indices of the ellipsometric observables $\Psi$ and $\Delta$}
	 	\textbf{a--b.} The first-order indices  $s^{\Psi,\Delta}_{\eps,\dperp}$ of  $\Psi(\theta,\omega)$ (a) and $\Delta(\theta,\omega)$ (b) for input variables $\eps\left(\omega\right)$ and $\dperp(\omega)$. 
            \textbf{c--d.} Angle-averaged first-order indices $\langle s^{\Psi,\Delta}_{\eps,\dperp}\rangle_\theta$. The background color gradient from blue to yellow indicates the increased $\dperp$ contributions towards higher frequencies.
            The frequency window \SIrange{2.1}{5.9}{\eV} in our experimental analysis is shown by the two vertical dashed lines.
            }
    \label{smfig:sobol_first}
\end{figure*}

In our Sobol$'$ analysis, we generate 40960 samples within a wide parameter space given by $\Re(\eps)\in \left[-50,0\right]$, $\Im(\eps)\in \left[0,10\right]$, $\Re(\dperp)\in \left[-3,3\right]\mathrm{nm}$, and $\Im(\dperp)\in \left[-3,3\right]\mathrm{nm}$. The result of the total-order indices is shown in Fig. 2 and the result of the first-order indices is presented in Fig.~\ref{smfig:sobol_first}. The first-order and total indices jointly indicate that it is possible to ellipsometrically measure $d$-parameters for high frequencies above the Drude frequency regime.

\subsection{The first-order partial differentiation analysis}
\label{smsec:diff_sensi}
\begin{figure*}[htbp]
    \centering
    \includegraphics[width=0.6\linewidth]{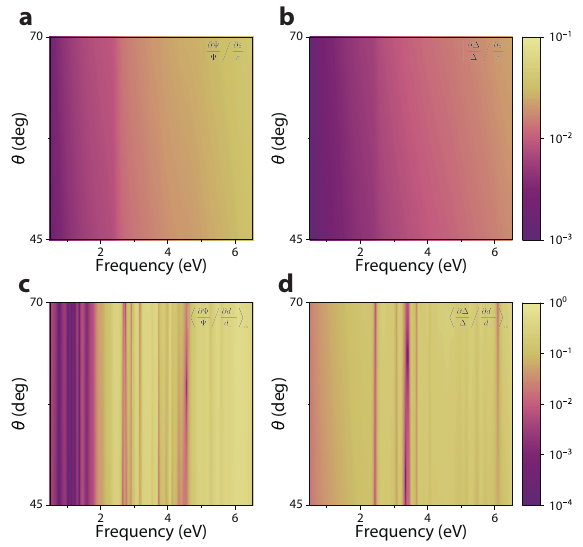}
    \caption{
            \textbf{Partial differentiation analysis.}
	 	\textbf{a--b.} Normalized first-order partial differentiation $\left|\frac{\partial \Psi}{\Psi}\bigg/\frac{\partial \eps}{\eps}\right|$ (a) and $\left|\frac{\partial \Delta}{\Delta}\bigg/\frac{\partial \eps}{\eps}\right|$ (b).
            \textbf{c--d.} Normalized first-order partial differentiation $\left|\left<\frac{\partial \Psi}{\Psi}\bigg/\frac{\partial \dperp}{\dperp}\right>_\alpha\right|$ (c) and $\left|\left<\frac{\partial \Delta}{\Delta}\bigg/\frac{\partial \dperp}{\dperp}\right>_\alpha\right|$ (d), where $\left<~\right>_\alpha$ denotes the average along all possible differential directions in the complex $\dperp$ plane.
          }
    \label{smfig:first_diff}
\end{figure*}

Complementary to the comprehensive Sobol$'$ analysis, we also use ordinary first-order partial differentiation to evaluate the contributions of various input parameters. 
We perform the differentiation numerically and the results are shown in Fig.~\ref{smfig:first_diff}. 
The permittivity $\eps$ adopts the tabulated data from Johnson \& Christy~\cite{johnson1972optical}. 
For the differentiation of $\dperp$, we perform averaging along all differentiation directions on the complex $\dperp$ plane because the evolution direction of $\dperp$ with regard to frequency is unknown (whereas this is known for the bulk permittivity). 

The differentiation result shows an abrupt increase of sensitivity to $\dperp$ for frequencies larger than around $\SI{2}{\eV}$, where the sensitivity to $\dperp$ becomes comparable with that to $\eps$. 
A representative value of $\dperp$ is chosen as $-1+0.5\iu$~nm in Fig.~\ref{smfig:first_diff}, and we have verified that the differentiation results in Fig.~\ref{smfig:first_diff} remains similar for other choices of $\dperp$.
This result thus reaches consistency with the Sobol$'$ analysis (Fig. 2 and Fig.~\ref{smfig:sobol_first}).

\section{Sample characterizations}
\label{smsec:sample_characterization}

Our samples are commercially available atomically flat template-stripped Au(111) chips (Platypus Technologies).
The thickness of the gold film is $\sim$\SI{100}{nm} and can be considered bulk for the frequency regime of interest.
We characterized the crystalline structure and the surface roughness of the gold chips.

\subsection{Crystalline structure}
\label{smsec:crystal}
\begin{figure}[htbp]
    \centering
    \includegraphics[width=0.6\linewidth]{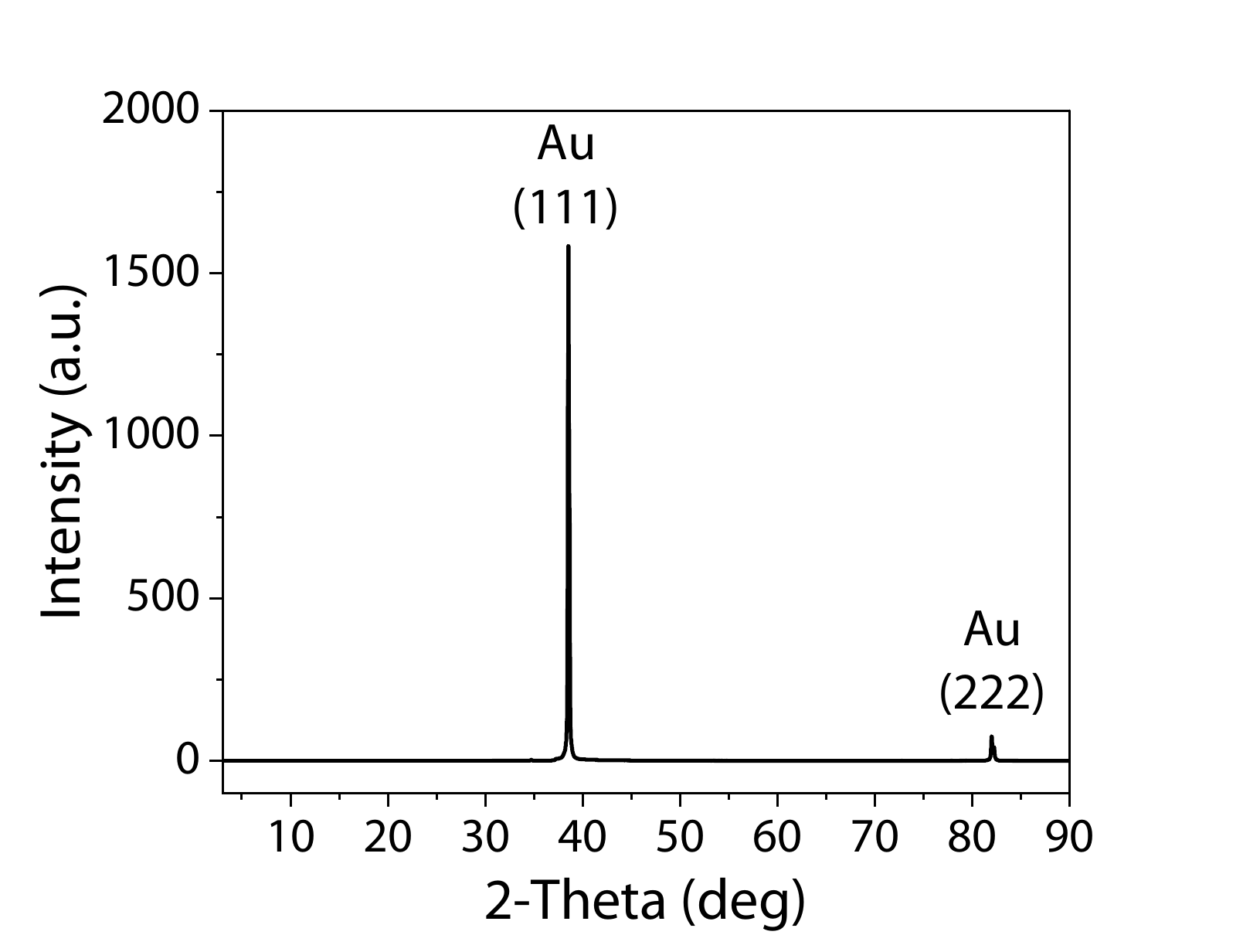}
    \caption{\textbf{X-ray diffraction analysis of the gold sample.} There exist two Bragg peaks corresponding to the $\mathrm{(111)}$ and $\mathrm{(222)}$ directions with the former peak in dominance.
    }
    \label{smfig:crystal}
\end{figure}

The crystalline structure of the gold films is dominated by the (111) phase. We confirm it by an X-ray diffraction (XRD) system (Rigaku Miniflex X-ray diffractometer).
Figure~\ref{smfig:crystal} presents the XRD patterns in the scanning range of \SI{3}{\degree}\,--\,\SI{90}{\degree}.
The intense peaks at $\SI{38.5}{\degree}$ and $\SI{82.0}{\degree}$ correspond to the $\mathrm{(111)}$ and $\mathrm{(222)}$ gold orientations, respectively, which confirms the dominance of the Au(111) phase.

\subsection{Surface roughness}
\label{smsec:surfacerough}
\begin{figure}[htbp]
    \centering
    \includegraphics[width=0.6\linewidth]{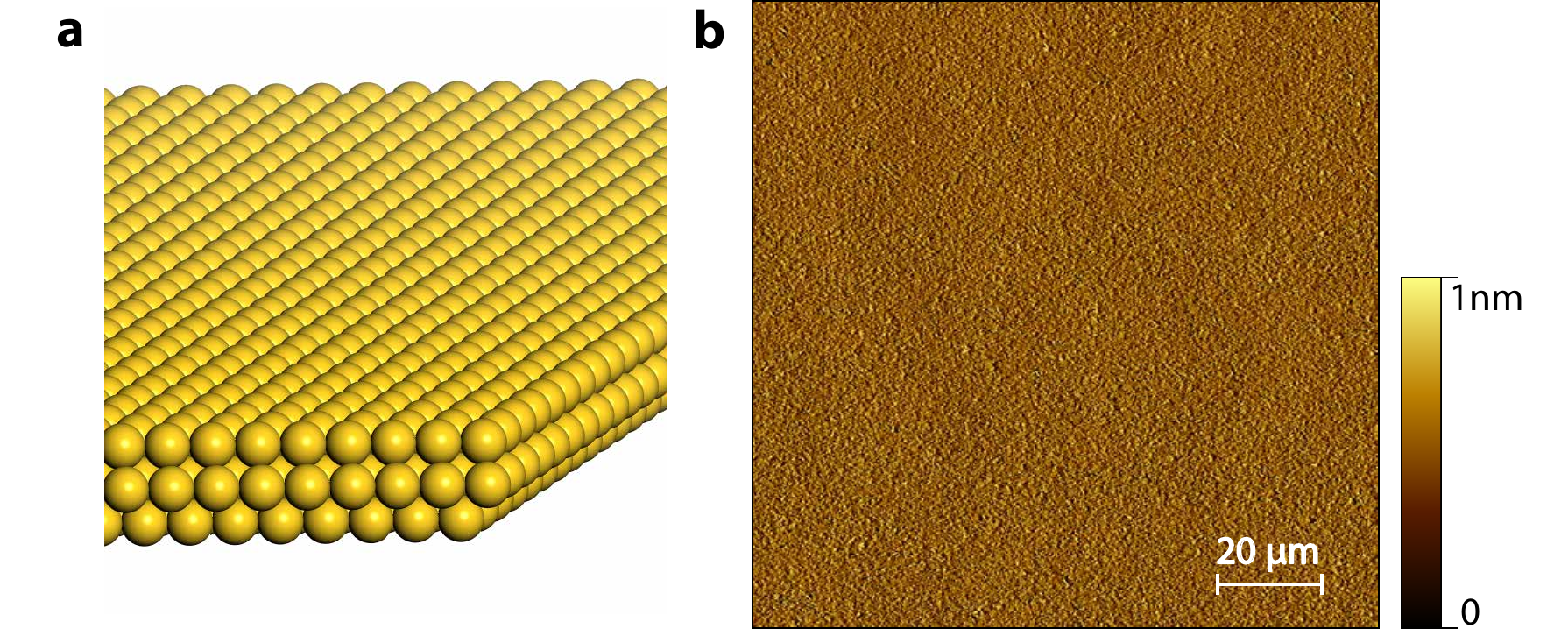}
    \caption{\textbf{Surface roughness.}  \textbf{a.}~Illustration of an idealized atomically flat Au(111) surface. \textbf{b.}~AFM image of the gold sample. The rms surface roughness is \SI{0.311}{nm}.
    }
    \label{smfig:surfacerough}
\end{figure}

Figure~\ref{smfig:surfacerough}a shows that the idealized flat Au(111) surface, which has an intrinsic roughness on the order of an atom size~\cite{yu1998atomic,namba2000modeling}. The surface roughness of the gold films is characterized using an atomic force microscope (MFP-3D Infinity AFM). The AFM image of our measured samples is shown in Fig.~\ref{smfig:surfacerough}b. The root-mean-square (rms) roughness of the gold films is \SI{0.311}{\nm}. 

\section{Ellipsometric measurements}
\label{smsec:ellipsometry}
Spectroscopic ellipsometry is widely used for the precise and non-destructive optical characterization of thin films.
It uses linearly polarized light to illuminate samples and detect the polarization-changed signals after interaction with them.
The change in polarization state is quantified by two raw measured parameters: amplitude ratio $\Psi$ and phase change $\Delta$, which is related by the reflectance ratio of the complex Fresnel reflection coefficients
for p- and s- polarized wave via $r_\mathrm{p} / r_\mathrm{s}  = \mathrm{ tan(\Psi) e^{ i \Delta } }$~\cite{tompkins2005handbook,fujiwara2007spectroscopic}.
In addition, ellipsometry is capable of measuring the reflectance of the p-polarized wave $R_{\rm{p}}$ and s-polarized wave $R_{\rm{s}}$ of the sample, which can be considered as two additional fitting variables.
Importantly, in the modulation and demodulation process of ellipsometry, $\Psi$ and $\Delta$ can be obtained from AC signal only, whereas the p- and s- intensity values contain both DC and AC contributions and are thus more vulnerable to external noises (such as the ambient light and $1/f$ noises).
Therefore, the incorporation of $R_\mathrm{p}$ and $R_\mathrm{s}$ will result in increased uncertainty of the fitting~\cite{fujiwara2007spectroscopic}.
As a result, only $\Psi$ and $\Delta$ are included as observables in our fitting model.

We performed ellipsometric measurements using an RC2 variable-angle spectroscopic ellipsometer. The wavelength range of the measurment spans from $\lambda = \SI{210}{\nm}$ to $\lambda = \SI{1690}{\nm}$.
The light incident angle of all measurements range from $\theta = \SI{45}{\degree} $ to $\theta = \SI{70}{\degree}$ with a step size of $\SI{0.2}{\degree}$. 
The fine angular resolution here is critical for our measurement because it provides sufficient observations for extracting the $d_\perp$ with unprecedented accuracy in previously inaccessible frequency regimes.
To validate the consistency of our data, we performed repeated measurements on five gold chips labeled as ``Au-1'' to ``Au-5'' at a fixed time interval of \SI{20}{\min} between neighboring measurements. 

For the ellipsometric measurement of each sample, measurements were performed within 4 hours after peel-off from the wafer to minimize environmental contamination.
The duration of angle scanning of the ellipsometer is \SI{20}{\min}, and each round of measurement was performed consecutively.

\begin{figure}[htbp]
    \centering
    \includegraphics[width=0.8\linewidth]{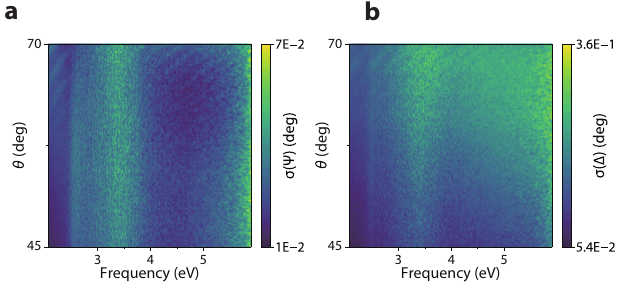}
    \caption{\textbf{Noise level in ellipsometry observables $\Psi$ and $\Delta$.}
    The standard deviation $\sigma(\Psi)$ and $\sigma(\Delta)$ of all ellipsometric measurements as a function of the incident angle $\theta$ and frequency $\omega$.
    }
    \label{smfig:StatisticDeviation}
\end{figure}

Figure~\ref{smfig:StatisticDeviation} shows the standard deviations of the total thirty ellipsometric measurements. The small standard deviations of $\Psi$ and $\Delta$ indicate the measurement results are clustered tightly around their mean values (See Fig. 1c and d), which are thus used for extracting the bulk permittivity and $d$-parameters in Fig. 3b and c.

\section{Fitting methodology}
\label{smsec:fit}

\subsection{Weighted nonlinear least-squares fitting}
\label{smsec:fit_description}
We adopt the nonlinear least-squares method to extract the bulk permittivity and $d$-parameters simultaneously from the measured $\Psi$ and $\Delta$.
We assume the $d$-parameters are independent of the incident angles (\ie neglecting the nonlocality of $d$-parameters), and thus the fitting is performed for each frequency at all angles with the loss function $L$ given by
\begin{equation}
    L = \sum_{i}^{N} \frac{{\left [ \Delta_{i}^{\mathrm{mes}}-\Delta_{i}^{\mathrm{mod}}\right ]}^{2}}{\sigma_{\Delta,i}^{2}}+\sum_{i}^{N} \frac{{\left [ \Psi_{i}^{\mathrm{mes}}-\Psi_{i}^{\mathrm{mod}}\right ]}^{2}}{\sigma_{\Psi,i}^{2}} ,
    \label{smeq:lossfunction}
\end{equation}
where $\Delta_{i}^{\mathrm{mes}}$ and $\Psi_{i}^{\mathrm{mes}}$ are measured observables, $\Delta_{i}^{\mathrm{mod}}$ and $\Psi_{i}^{\mathrm{mod}}$ are the model predictions. The summation is over all reflection angles indexed by $i$, and the residuals are weighted by the standard deviations of the experimental data $\sigma_{\Delta,i}$ and $\sigma_{\Psi,i}$.
Because the model is nonlinear, the Levenberg--Marquardt algorithm is used to solve the minimization problem. The confidence intervals are calculated in an iterative way based on the F-test.

In the fitting, we use the Johnson and Christy data of gold~\cite{johnson1972optical} as the initial guess for the bulk permittivity.
It is noted that by imposing $d$-parameters to be zero, the fitting process will produce the classical bulk permittivity as the standard ellipsometry does.
The initial value of $\dperp$ is randomly chosen with a magnitude of a few angstroms, and the fitting is repeated for several different initial $\dperp$ to avoid the trapping of the fitting at local optima. We have also assumed $\dpar=0$ in the fitting (see discussion and the relaxation of this assumption in Sec.~\ref{smsec:dpara}).

\subsection{Anterior-measurement Monte Carlo analysis}
\label{smsec:fit_method}
The previous supplementary section~\ref{smsec:model_sensi} suggests that $\dperp$ can be extracted from the quantum Fresnel equations by ellipsometric means. 
In the following, we show that $\dperp$ can be obtained through the weighted non-linear least-squares fitting from the measured observables under the experimental noise levels.
We follow the spirit of the Monte-Carlo method and ensemble statistics to support the fitting approach. The split-step process is described below.
\begin{enumerate}
    \item The parameter space is chosen as $\omega \in \left[2.1,5.9\right]\mathrm{eV}$, $\Re(\dperp
    ) \in \left[-2,2\right]\mathrm{nm}$, $\Im(\dperp
    ) \in \left[-1,3\right]\mathrm{nm}$. 
    $\eps\left(\omega\right)$ uses the measured dispersive permittivity. The large parameter space ensures that all possible $\dperp$ can be covered.
    Before conducting the measurement, the ground truths are randomly chosen without being associated with any particular interfaces such that we can show the general utility of the ellipsometric approach. 
    \item We generate $10^6$ ensembles at each noise level, where the parameter combinations are chosen randomly.
    \item Each ensemble generates ``clean'' observables $\Psi$ and $\Delta$ at 126 angles from $\SI{45}{\degree}$ to $\SI{70}{\degree}$ with a step size of \SI{0.2}{\degree}, as in the experimental conditions. Then, we add unbiased, normally distributed noise to create the ``contaminated'' observables at a certain noise level. To better mimic the experimental conditions, the standard deviations of $\Psi$ and $\Delta$ are correlated---the former is fixed at a quarter of the latter, which is consistent with their respective value ranges, \ie $\psi\in[24.9,45.0]\deg$ and $\Delta\in[70.5,173.5]$ .
    This is supported by the experimental conditions (Fig.~\ref{smfig:StatisticDeviation}) where the noise level of $\Psi$ varies from $\SI{0.012}{\degree}$ to $\SI{0.074}{\degree}$ while that of $\Delta$ from $\SI{0.054}{\degree}$ to $\SI{0.36}{\degree}$, forming an approximate factor of four difference (Fig. 2e).
    \item We apply the weighted non-linear least-square method to extract the bulk permittivity and $\dperp$ simultaneously from the observables. Then, we analyze the errors between the ground truths and the fitted values across the $10^6$ ensembles, which is shown in Fig. 2g--h in the main text.
    \item We generate another $10^4$ ensembles at each points in Fig. 2f. In that particular calculation, the assumed correlation between $\sigma(\Psi)$ and $\sigma(\Delta)$ is relaxed, and the rest of the steps remain unchanged. 
    
\end{enumerate}

\subsection{Posterior-measurement Monte Carlo analysis}
\label{smsec:posterior}

\begin{figure}[htbp]
    \centering
    \includegraphics[width=\linewidth]{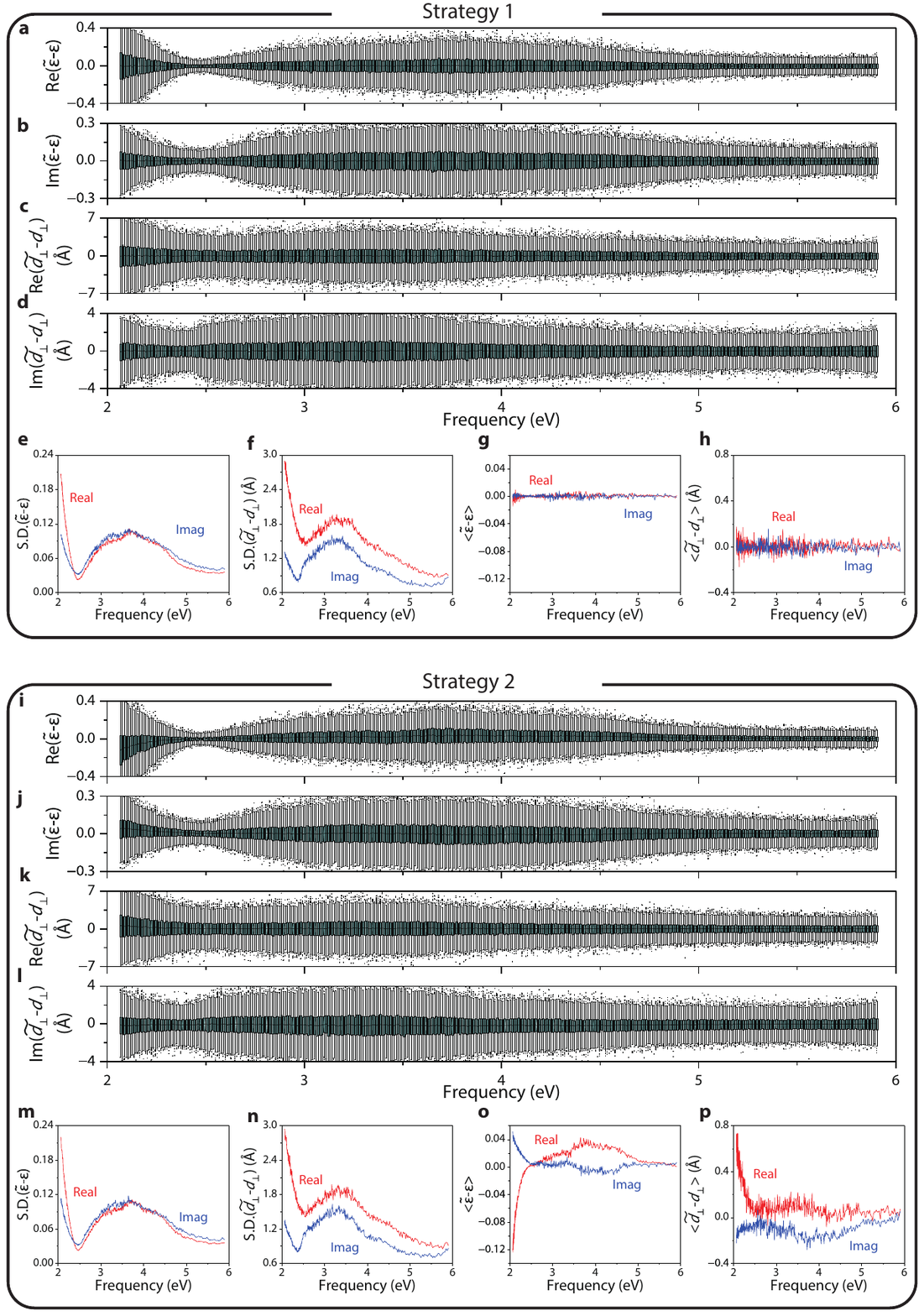}
    \caption{\textbf{Error analysis of posterior Monte-Carlo simulations.}
        \textbf{a--d.} Box plots for fitting errors of strategy 1. Black lines indicate the mean values, boxes are from the first quartile to the third quartile, whiskers are in 1.5 times the interquartile range, and discrete points are outliers.
        \textbf{e--h.} Statistics of $10^3$ ensembles at each frequency in strategy 1: standard deviation of fitting errors for $\eps$ (e) and $\dperp$ (f), and mean errors of $\eps$ (g) and $\dperp$ (h).
        \textbf{i--p.} Same as a--h but for Strategy 2. 
        The standard deviations of errors are comparable for both Strategy 1 and Strategy 2; crucially, the estimates are unbiased and biased in Strategy 1 and Strategy 2, respectively (compare g--h with o--p). 
    }
    \label{smfig:posterior}
\end{figure}

The repeated measurements across samples and separated time (see \sisec{smsec:time_and_samples}) enlarge the database and enable the reduction of fitting errors.
We perform additional post-measurement Monte Carlo analysis, where the ground truths are chosen to be the experimentally extracted bulk permitivities and $d$-parameters (Fig. 3). By doing so, we aim to obtain more comprehensive noise features in our measurement and fitting.

The major goal here is to choose between two fitting strategies: 
\begin{enumerate}
    \item Strategy 1 is to first take arithmetic mean of repeated observables and then perform the fitting.
    \item Strategy 2 is to perform the fitting first using each set of individual observables and then take the arithmetic mean on the fitted results. 
\end{enumerate}
The two strategies are equivalent for linear fittings but may differ for nonlinear ones, which is the case of the quantum Fresnel equation [Eq. (2)].
Analogous to~\sisec{smsec:fit_method}, we utilize the Monte-Carlo method to compare the performance of two strategies based on the following steps.
\begin{itemize}
    \item We adopt the dispersive measured results obtained in Fig. 3 as the ground truth. For each frequency, we generate $10^3$ ensembles and each ensemble contains 30 samples as in our experiment (see \sisec{smsec:sample}).
    \item Each sample generates ``clean'' observables $\Psi$ and $\Delta$ at 126 angles from $\SI{45}{\degree}$ to $\SI{70}{\degree}$ with a step size of \SI{0.2}{\degree}, matching experimental conditions. Then, we add unbiased noises to create the ``contaminated'' observables in which the noises are normally distributed, obeying the statistics in Fig.~\ref{smfig:StatisticDeviation} at each angle for each frequency.
    \item For both strategies, we apply the weighted non-linear least-square method (\sisec{smsec:fit_description}) to extract the bulk permittivity and $\dperp$ simultaneously from the observables. 
    For Strategy 1, we generate 30 samples (the same as the number of repeated measurements in our experiment), average them, and perform the fitting with the averaged samples to generate the result within each ensemble. 
    For Strategy 2, we also generate $10^3$ ensembles (30 samples within each), perform the fitting individually, and perform statistics to evaluate the arithmetic mean and errors within each ensemble. 
\end{itemize}

The comparison between the two strategies is shown in Fig.~\ref{smfig:posterior}, where the top box corresponds to Strategy 1 and the bottom box to Strategy 2.
Fig.~\ref{smfig:posterior}a--d and i--l show the box plots of the errors (estimates subtracted by ground truths) in the Monte Carlo simulations. Notably, the uncertainty is minimal in the frequency regime where the Bennett mode appears (see the ``waist'' of uncertainty near \SI{2.5}{\eV}), which is consistent with the most pronounced residual reduction in Fig. 3a.
Fig~\ref{smfig:posterior}e, f, m, and n show that the uncertainty of the two strategies are comparable. 
It is worth noting that the standard deviation here refers to that of the mean of 30 samples in each ensemble (such that they are directly comparable to the fitting uncertainty Fig. 3b and c), whereas that in Fig. 2 refer to the standard deviation of a single measurement. Consequently, the former is smaller and the two types of standard deviations are related to each other by a factor of $\sqrt{N}$, where $N=30$ is the number of repeated measurements.

Fig.~\ref{smfig:posterior}g, h, o, and p compare the average errors (without taking absolute values) of the two strategies. Importantly, Strategy 1 is unbiased (Fig.~\ref{smfig:posterior}g and h); however, Strategy 2 produces biased estimates (Fig.~\ref{smfig:posterior}o and p) despite the noise added containing zero bias.   
As a result, we adopt Strategy 1 in our experimental data processing: we take the average of $\Psi$ and $\Delta$ (Fig. 1c and d) in the 30 experimental measurements (\sisec{smsec:time_and_samples}) first and then perform the fitting to extract $\eps$ and $\dperp$ simultaneously.

\begin{figure}[htbp]
    \centering
    \includegraphics[width=1\linewidth]{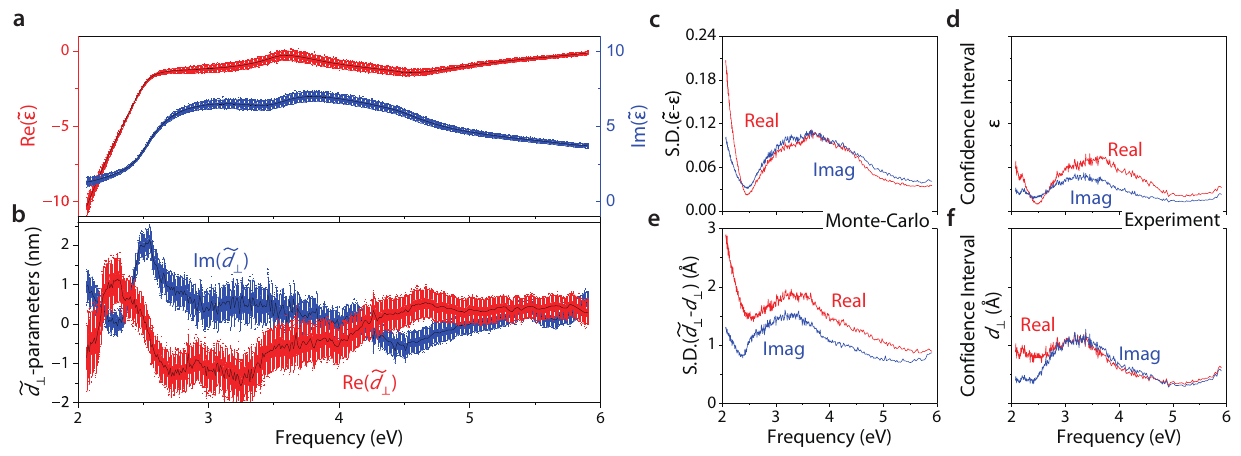}
    \caption{\textbf{Comparsion between posterior Monte-Carlo simulations with experimental data.}
    \textbf{a.} Box plot of real (red) and imaginary (blue) parts of Monte Carlo estimates $\tilde{\eps}$.
    \textbf{b.} Box plot of real (red) and imaginary (blue) parts of Monte Carlo estimates $\tilde{\dperp}$.
    Black lines indicate the mean values, boxes are the standard deviation $\pm1\sigma$, whiskers denote the data in the range of $\pm3\sigma$, and the discrete points are the outliers.
    \textbf{c--d.} Monte Carlo standard deviations of $\eps$ (c) and experimental fitting confidence interval of $\eps$ (d), both corresponding to $1\sigma$.
    \textbf{e--f.} Same as c and d but for $\dperp$.
    }
    \label{smfig:posterior_add_truth}
\end{figure}

For completeness, we also plot the posterior Monte Carlo estimates without subtracting the ground truth in Fig.~\ref{smfig:posterior_add_truth}a and b for direct comparison with the experimental fitting in Fig. 3b and c. 
Although the experimental confidence intervals are slightly smaller in magnitude (which may be due to cross-measurement correlations in the experiment), the general lineshapes of the Monte Carlo frequency-dependent uncertainty (Fig.~\ref{smfig:posterior_add_truth}c and e) are accordant with those of the experimental confidence intervals (Fig.~\ref{smfig:posterior_add_truth}d and f).

\vspace{0.2cm}
Taken together, Secs.~\ref{smsec:fit_description}, \ref{smsec:fit_method}, and \ref{smsec:posterior} jointly confirm the feasibility of extracting bulk permittivities and $d$-parameters using angle-resolved ellipsometry and judiciously designed fitting strategy. 

\section{Temporal and cross-sample consistency}
\label{smsec:time_and_samples}
\subsection{Temporal consistency}
\label{smsec:time}

\begin{figure}[htbp]
    \centering
    \includegraphics[width=0.8\linewidth]{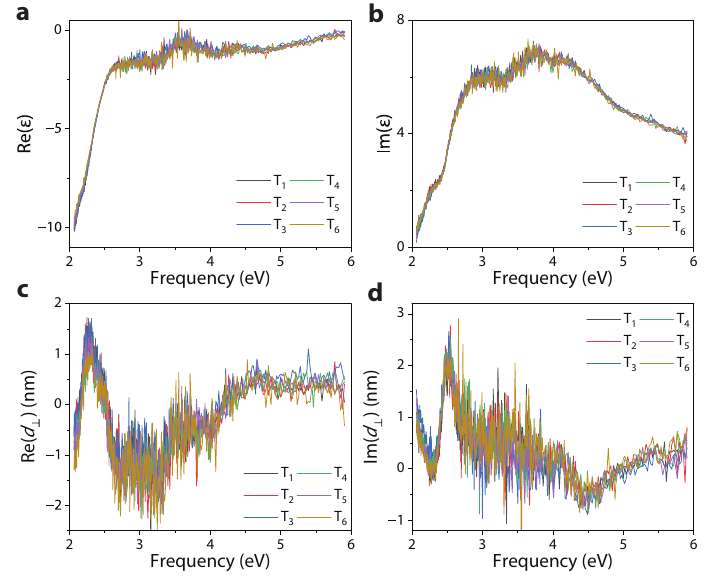}
    \caption{\textbf{Temporal consistency in the measured permittivity and $d$-parameters.}
    \textbf{a-d.} Extracted $\Re{\eps(\omega)}$ (a), $\Im{\eps(\omega)}$ (b), $\Re{\dperp(\omega)}$ (c), and $\Im{\dperp(\omega)}$ (d) at different time points.
    }
    \label{smfig:time_consistency}
\end{figure}

To investigate the temporal consistency of our measurement, we compared the cross-sample averaged ellipsometric results at all time points (i.e., $\mathrm{T}_1$ to $\mathrm{T}_6$) separated by a fixed time interval of 20 minutes between adjacent measurements.
The fitting results from $\mathrm{T}_1$ to $\mathrm{T}_6$ exhibit a high level of consistency, as shown in Fig.~\ref{smfig:time_consistency}. 

\subsection{Cross-sample consistency}
\label{smsec:sample}

\begin{figure}[htbp]
    \centering
    \includegraphics[width=0.8\linewidth]{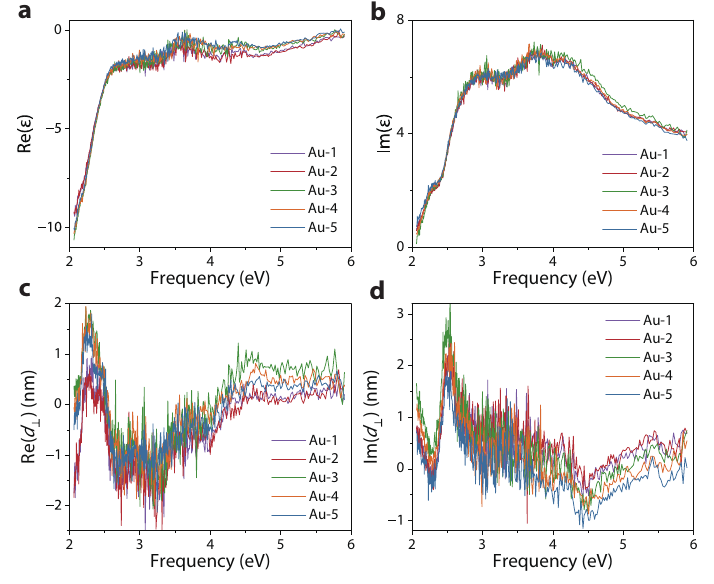}
    \caption{\textbf{Cross-sample consistency in the measured permittivity and $d$-parameters.}
    \textbf{a-d.} Extracted $\Re{\eps(\omega)}$ (a), $\Im{\eps(\omega)}$ (b), $\Re{\dperp(\omega)}$ (c), and $\Im{\dperp(\omega)}$ (d) from different gold samples.
    }
    \label{smfig:sample_consistency}
\end{figure}

Next, we average the ellipsometric measurement results for each sample at different time points to study the consistency among samples Au-1 to Au-5.
Figure~\ref{smfig:sample_consistency} presents the time average fitting results of $\eps$ and $\dperp$ for the five gold samples, which again exhibit high consistency. Overall, the fitting results confirm the cross-sample stability of the material properties in our measurements.

Jointly shown in Figs.~\ref{smfig:time_consistency} and~\ref{smfig:sample_consistency}, the consistency of our data across different samples and measurement times indicates the robustness and stability of our experiment and samples. Such consistency also justifies the use of the entire dataset for data analysis.

\section{Alternative fitting on $\dperp-\dpar$}
\label{smsec:dpara}
In the fitting process described in the main text, we adopt the assumption $\dpar=0$ that is generally expected for charge-neutral interfaces. However, such assumption may be violated at some frequencies under specific crystalline structures like the $\mathrm{(111)}$ noble-metal surfaces (as in our case)
due to the potential contributions from the Shockley surface states~\cite{feibelman1982surface,echarri2021optical}. 

\begin{figure}[htbp]
    \centering
    \includegraphics[width=0.9\linewidth]{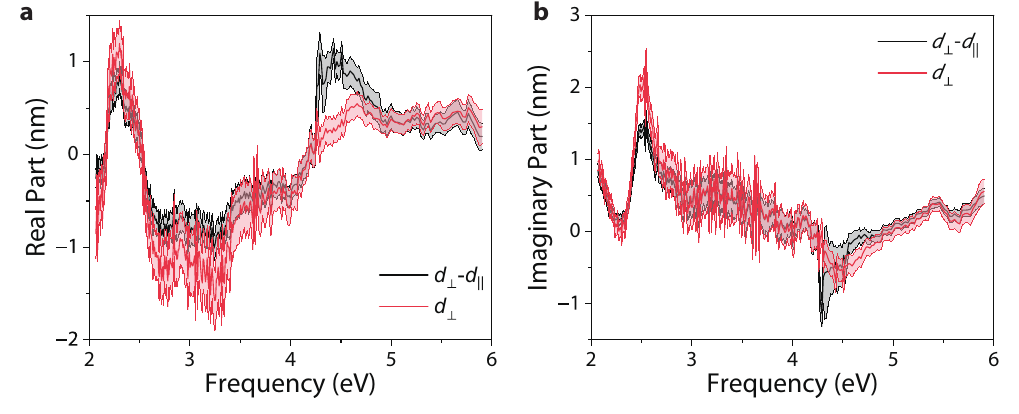}
    \caption{
            \textbf{Comparison between fitted $\dperp$ and $\dperp-\dpar$ with and without assuming $\dpar=0$.} Shading denotes the confidence interval of $3\sigma$ in each fitting.
    }
    \label{smfig:dpara}
\end{figure}

Bearing this in mind, we perform an alternative fitting on $\dperp-\dpar$ based on a first-order expansion of the quantum Fresnel law. The ratio of the quantum Fresnel reflection coefficients $\rho=r_\mathrm{p}/r_\mathrm{s}$ can be expressed as~\cite{liebsch2013electronic}
\begin{equation}
    \rho = \rho^{\mathrm{cl}}\left[1+2\iu k_{z,\mathrm{1}}\left(\dperp-\dpar \right)\frac{\eps}{\eps \cot^\mathrm{2}\theta-1} \right],
    \label{smeq:rho}
\end{equation}
where $\rho^{\mathrm{cl}}=r^\mathrm{cl}_\mathrm{p}/r^\mathrm{cl}_\mathrm{s}$ is the ratio from classical Fresnel law. We adopt the same fitting method (described in Section~\ref{smsec:fit_description}) to this model and obtain the fitted results on $\dperp-\dpar$ shown in Fig.~\ref{smfig:dpara}. 
The fitted $\dperp-\dpar$ (using Eq.~\eqref{smeq:rho} without assuming $\dpar=0$) and $\dperp$ (using Eq. (2) and assuming $\dpar=0$) agree with each other quantitatively.

\section{Kramers--Kronig extrapolation}
\label{smsec:applyKK}

In the Kramers--Kronig relation Eq.~\eqref{smeq:KK}, the integration is performed from zero (DC) to infinite frequency. However, practical measurements are always performed under finite frequency windows, which thus requires extrapolation for performing the Kramers--Kronig transformation. 

\begin{figure}[htbp]
    \centering
    \includegraphics[width=0.6\linewidth]{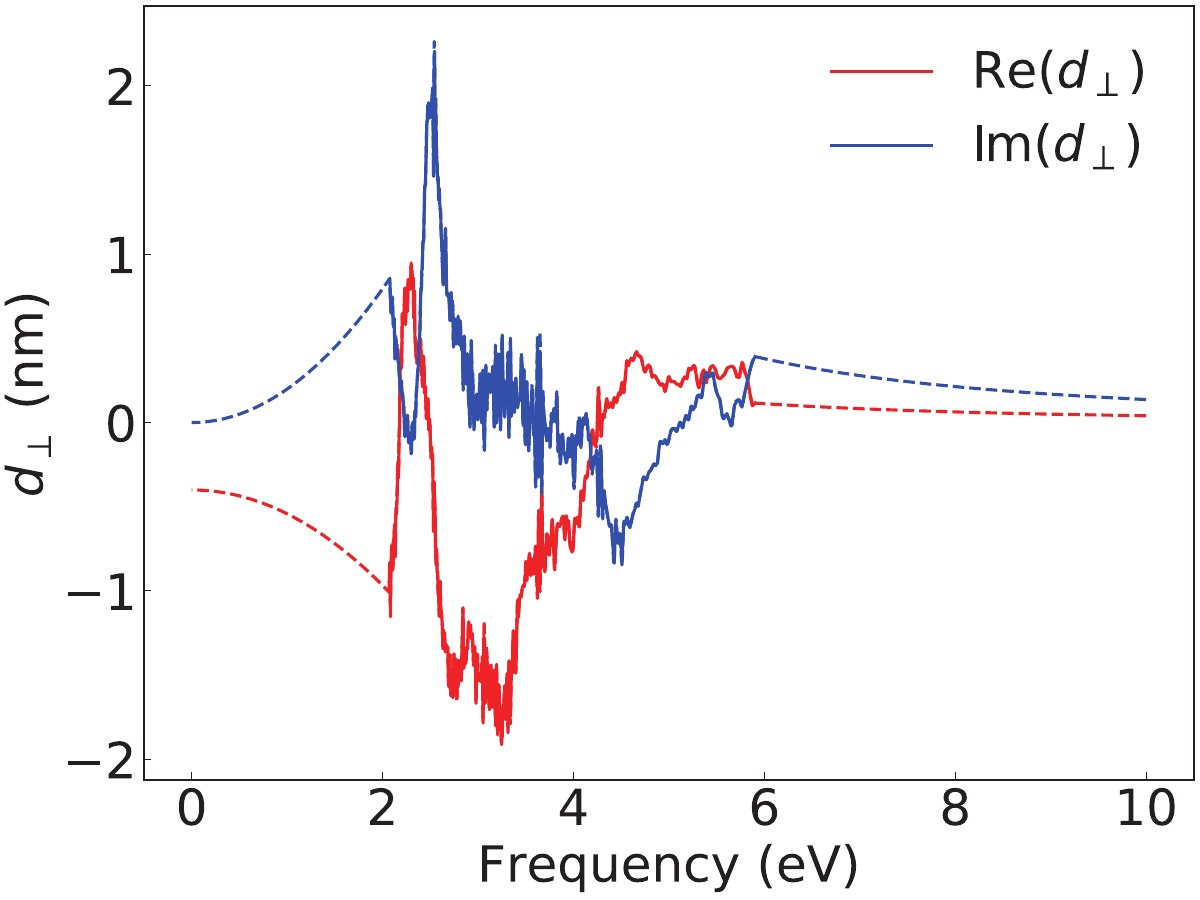}
    \caption{
            \textbf{Data used for Kramers--Kronig transformation.}
            Measured data is shown in solid lines, while the extrapolated quadratic decays toward zero and infinite frequencies are shown in dashed lines. The asymptotic limits are \SI{0}{\nm} for $\omega\to\infty$ and \SI{-0.4}{\nm} for $\omega\to 0$, respectively.
            }
    \label{smfig:extrapolation}
\end{figure}

We perform extrapolation in the following manner. First, toward higher frequencies, it has been derived~\cite{persson1983sum} that $\dperp\left(\omega\right)\sim \omega^{-2}$ as $\omega\to\infty$. We thus impose this condition together with the continuous condition at the high-frequency cutoff. Second, toward DC frequencies below the (screened) plasma frequency, it has been shown that $\Re{\dperp}$ tends to a constant while $\Im{\dperp}$ vanishes gradually (see \eg Ref.~\citenum{liebsch2013electronic} and other calculations of $d$-parameters). Therefore, we impose quadratic decays to $\SI{-0.4 }{\nm}$ and $\SI{0}{\nm}$ for $\Re{\dperp(\omega)}$ and $\Im{\dperp(\omega)}$, respectively, where the asymptotic values are obtained from a previous measurement in the near-infrared regime~\cite{yang2019general}. The measurement data and their extrapolation are shown in Fig.~\ref{smfig:extrapolation}. 

We further note that only KK transformation close to the cut-off of the measurement frequency window will be affected by the particular choice of extrapolation. Within the measurement frequency window, the choice of extrapolation has a very limited impact on the transformation because of the quadratic denominator $\omega^{'2}-\omega^2$ in Eq.~\eqref{smeq:KK}.

\section{Energy dissipation due to the surface dipole}
\label{smsec:energydis}
By virtue of energy conservation, the energy of the incident wave impinged on a planar surface can be divided into three parts: reflection, transmission, and surface dissipation. 
However, the classical electrodynamic view ignores the surface dissipation term, leading to the well-known energy conservation between reflection and transmission. 
One can obtain the surface energy dissipation by comparing the classical and quantum Fresnel reflection and transmission coefficients. classically, the energy conservation law is 
\begin{equation}\label{smeq:power_c}
    P_{\mathrm{inc}}=P_{\mathrm{ref}}+P_{\mathrm{tra}} = RP_{\mathrm{inc}}+TP_{\mathrm{inc}},
\end{equation}
where $P_\mathrm{inc}$, $P_\mathrm{ref}$, and $P_\mathrm{tra}$ are the incident, reflected, and transmitted power, respectively, and $R$ and $T$ are power reflection and transmission coefficients, respectively. Evidently, $R+T=1$.

Taking into account the contribution of $d$-parameters, there is extra energy dissipation on the surface. The energy conservation law is thus modified as 
\begin{equation}\label{smeq:power_nc}
    P_{\mathrm{inc}}=P^{\mathrm{nc}}_{\mathrm{ref}}+P^{\mathrm{nc}}_{\mathrm{tra}}+P_{\mathrm{surface}} = R^{\mathrm{nc}}P_{\mathrm{inc}}+T^{\mathrm{nc}}P_{\mathrm{inc}}+\eta P_{\mathrm{inc}}.
\end{equation}
where the superscript ``nc'' stands for ``nonclassical'', and the surface energy dissipation ratio is $\eta$. 
Combining Eqs.~\eqref{smeq:power_c} and \eqref{smeq:power_nc}, we can obtain
\begin{equation}
    \eta = 1-R^{\mathrm{nc}}-T^{\mathrm{nc}}.
    \label{smeq:surface_dissipation}
\end{equation}

\begin{figure}[htbp]
    \centering
    \includegraphics[width=0.5\linewidth]{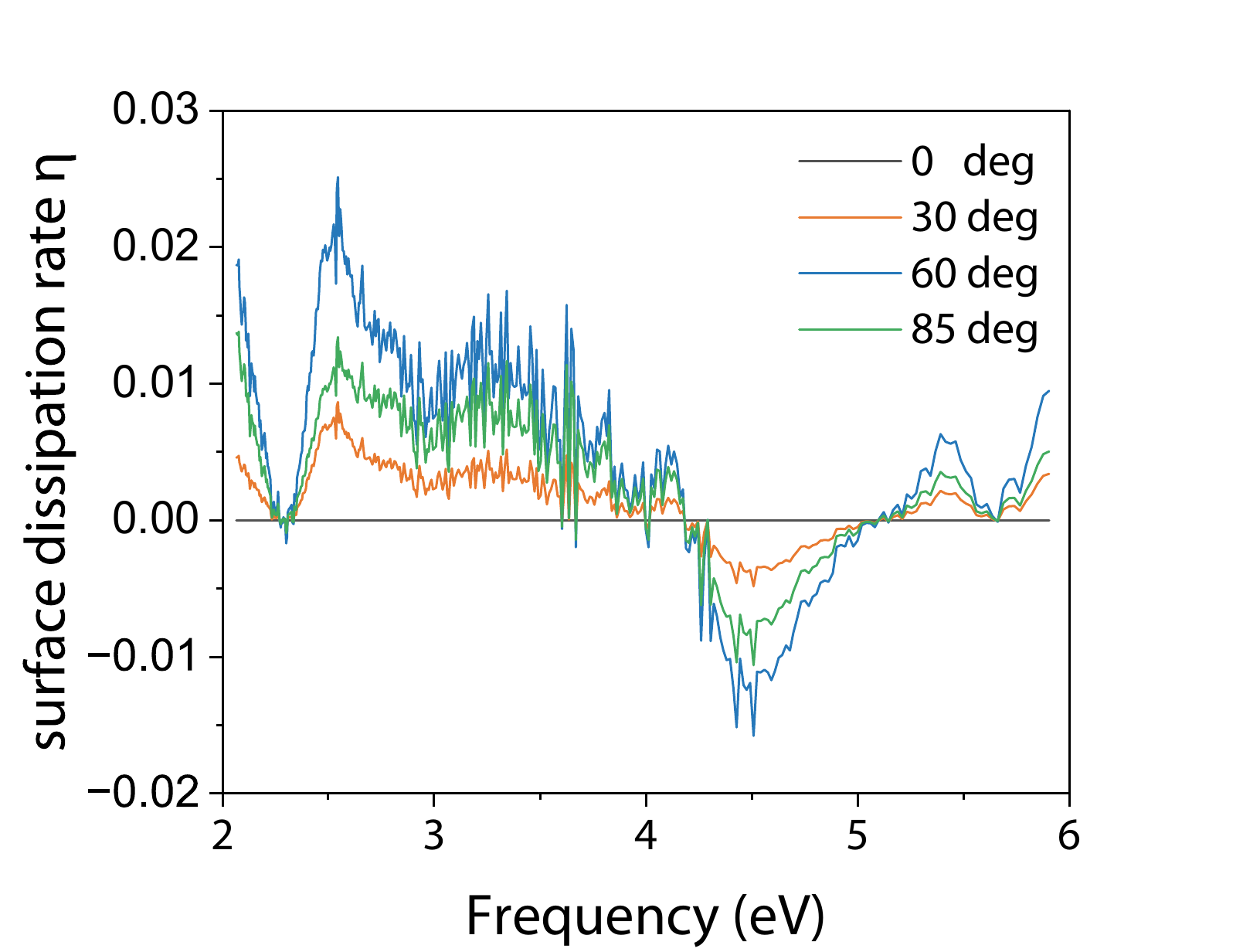}
    \caption{\textbf{Nonclassical Surface energy dissipation ratio at different incident angles.}}
    \label{smfig:enter-label}
\end{figure}

Based on Eq.~\eqref{smeq:surface_dissipation}, we apply the measured $d$-parameters to obtain the surface energy dissipation at different incident angles. 
At normal incidence, there is no surface dissipation, which increases for larger incident angles, becomes maximal at an intermediate angle, and decreases again for grazing angles. 
That is because during the increasing of incident angle from \SI{0}{\degree} to \SI{90}{\degree}, the impact of $\dperp$ will first increase with the scaling of $q^2\dperp$. With the incident angle approaching \SI{90}{\degree}, the classical result is restored---transmission vanishes, and reflection tends to unity.
The lineshapes of surface dissipation at nonzero and non-90-degree incident angles are qualitatively similar to $\Im\dperp$. 
{Consistent with the passivity evaluation in Fig. 3d in the main text, the existence of the likely-spurious negative surface dissipation near \SI{4.5}{\eV} is noticed, which may be due to imperfection in the experimental conditions or underestimated $\dpar$ contributions (notably, the difference between two fitted results (Fig.~\ref{smfig:dpara}) with and without assuming $\dpar=0$ is also the largest near \SI{4.5}{\eV}, suggesting that there may be considerable $\dpar$ at those frequencies).

\section{Measurement limitation}
\label{smsec:limit}

Despite the analytical, experimental, and statistical efforts described above, our measurements still have their limitations. For clarity, completeness, and potential future improvement, we discuss the limitations below:
\begin{enumerate}
    \item \textit{Surface adsorption contamination.} The samples were measured once peeled from the wafer, and the total measured time was controlled within 4 hours. The measurement time could not be further compressed due to the alignment and angle scanning in each round. During this process under ambient conditions, avoiding undesired surface absorption can be difficult. For example, the adsorbed airborne carbonaceous will decrease the hydrophilicity of the gold surface, thus possibly affecting the measurement~\cite{olmon2012optical,lee1996adsorption,chai2007large}. 
    To mitigate this limitation in the future, it may be possible to perform sample preparation and ellipsometry in vacuum conditions for which products have been developed, \eg using a VUV-VASE ellipsometer~\cite{hilfiker2000optical} to isolate the atmosphere from pure dry nitrogen gas thus reduce contamination.
    \item \textit{Surface charges.} When the gold samples were peeled from the silicon wafer, there might be charges generated at both surfaces due to friction. These possible charges may invalidate the charge-neutral surface assumption, leading to a nonzero $\dpar$.
    \item \textit{Surface roughness.} Although our sample is already atomically flat with roughness rms $\approx \SI{0.3}{\nm}$, its magnitude is still comparable with the measured $d$-parameters in certain frequency regimes. It is thus pertinent to advance sample preparation such that the surface roughness can approach the natural crystallinity limit (\eg for Au(111) the surface roughness should be on the order of $\SI{1}{\angstrom}$).
\end{enumerate}

\section[Quantum corrections to LDOS and EELS]{Quantum corrections to local density of states and electron energy loss}
\label{smsec:quantumenhance}
In this section, we briefly summarize the analytical approaches for calculating the local density of states (LDOS) and electron energy loss with $d$-parameter contributions taken into account. 
The additional calculations are shown in Fig.~\ref{smfig:enhancement}, which complement Fig. 4a and b in the main text.

\subsection{LDOS near a planar surface}\label{smsec:ldos_plane}
The local density of states of a point dipole at a distance $h$ away from a $z=0$ surface (material 1 above and material 2 below, same as the main text convention) can be partitioned into perpendicular and parallel terms as~\cite{novotny2012principles,carminati2015electromagnetic,gonccalves2020plasmon} 
\begin{subequations}\label{smeq:ldos_plane}
    \begin{align}
        &\frac{\rho_\perp(\omega,h)}{\rho_0\left(\omega\right)}=1+\frac{3}{2}\Re\int_0^\infty \diff u\, \frac{u^3}{\sqrt{1-u^2}}r_\mathrm{p}~\e^{2\iu k_1 h \sqrt{1-u^2}},\\
        &\frac{\rho_\parallel(\omega,h)}{\rho_0\left(\omega\right)}=1+\frac{3}{4}\Re\int_0^\infty \diff u\, \frac{u}{\sqrt{1-u^2}}\left[r_\mathrm{s}-(1-u^2)r_\mathrm{p}\right] \e^{2\iu k_1 h \sqrt{1-u^2}},
    \end{align}
\end{subequations}
where $u=q/k_1$, $k_1$ is the wavevector in material 1, $\rho_0$, $\rho_\perp$, and $\rho_\parallel$ denote the original, enhanced perpendicular and parallel LDOS, respectively. 
Evidently, the nonclassical correction enters via the modified reflection coefficients.
For randomly oriented dipoles, the orientation-averaged LDOS is $\langle\rho\rangle=\frac{1}{3}\rho_\perp+\frac{2}{3}\rho_\parallel$. 

\subsection{LDOS near a sphere}
\label{smsec:ldos_sphere}
The LDOS of a dipole emitter located near a nanosphere is given by~\cite{chew1987transition,christensen2014nonlocal}
\begin{subequations}\label{smeq:ldos_sphere}
    \begin{align}
        &\frac{\rho_\perp(\omega,h)}{\rho_0\left(\omega\right)}=1+\frac{3}{2}\frac{1}{y^2}\sum_{l=1}^\infty(2l+1)l(l+1)\Re\left\{ -a_{\mathrm{p},l}\left[h_l^{(1)}(y)\right]^2\right\},\\
        &\frac{\rho_\parallel(\omega,h)}{\rho_0\left(\omega\right)}=1+\frac{3}{4}\frac{1}{y^2}\sum_{l=1}^\infty(2l+1)\Re\left\{ -a_{\mathrm{p},l}\left[\xi'_l(y)\right]^2-a_{\mathrm{s},l}\left[\xi_l(y)\right]^2\right\},
    \end{align}
\end{subequations}
where $\rho_\perp(\omega,h)$ and $\rho_\parallel(\omega,h)$ correspond to the LDOS extracted from vertically and horizontally oriented (with regard to the closest spherical surface) dipoles, respectively, $y=k_1(R+h)$, $\xi_l(x)=xh_l^{(1)}(x)$, and $h_l^{(1)}$ is the spherical Hankel function of the first kind. $l$ is a positive integer. The prime superscript notation denotes the differentiation with respect to the arguments inside the parenthesis. $a_{\mathrm{p},l}$ and $a_{\mathrm{s},l}$ are the nonclassical Mie coefficients incorporating $d$-parameters~\cite{gonccalves2020plasmon}
\begin{subequations}\label{smeq:ldos_sphere_mie}
    \begin{align}
        &a_{\mathrm{p},l}=\frac{\eps_2 j_l(x_2)\Psi_l'(x_1)-\eps_1 j_l(x_1)\Psi_l'(x_2)+(\eps_2-\eps_1)\left\{j_l(x_1)j_l(x_2)\left[l(l+1)\right]\dperp+\Psi_l'(x_1)\Psi_l'(x_2)\dpar\right\}R^{-1}}{\eps_2 j_l(x_2)\xi_l'(x_1)-\eps_1 h_l^{(1)}(x_1)\Psi_l'(x_2)+(\eps_2-\eps_1)\left\{h_l^{(1)}(x_1)j_l(x_2)\left[l(l+1)\right]\dperp+\xi_l'(x_1)\Psi_l'(x_2)\dpar\right\}R^{-1}},\\
        &a_{\mathrm{s},l}=\frac{j_l(x_2)\Psi_l'(x_1)-j_l(x_1)\Psi_l'(x_2)+(x_2^2-x_1^2)j_l(x_1)j_l(x_2)\dpar R^{-1}}{j_l(x_2)\xi_l'(x_1)-h_l^{(1)}(x_1)\Psi_l'(x_2)+(x_2^2-x_1^2)h_l^{(1)}(x_1)j_l(x_2)\dpar R^{-1}},
    \end{align}
\end{subequations}
where $x_j = k_jR$, $j\in\{1,2\}$, $\Psi_l(x)=xj_l(x)$, and $j_l$ is the spherical Bessel function of the first kind.

First, we examine the average LDOS of a randomly oriented dipole emitter near a gold sphere~\cite{christensen2014nonlocal,gonccalves2020plasmon,eriksen2024nonlocal} close to the dipolar localized surface plasmon resonance in Fig. 4a (\sisec{smsec:ldos_sphere}), where \mbox{on-resonance} LDOS is further boosted with quantum corrections taken into account. 
Also incidentally, detuned from the dipolar resonance towards small frequencies around $\SI{2.1}{\eV}$, the classical reduction of LDOS is notably plateaued by the Landau damping.

\begin{figure}[htbp]
    \centering
    \includegraphics[width=0.8\linewidth]{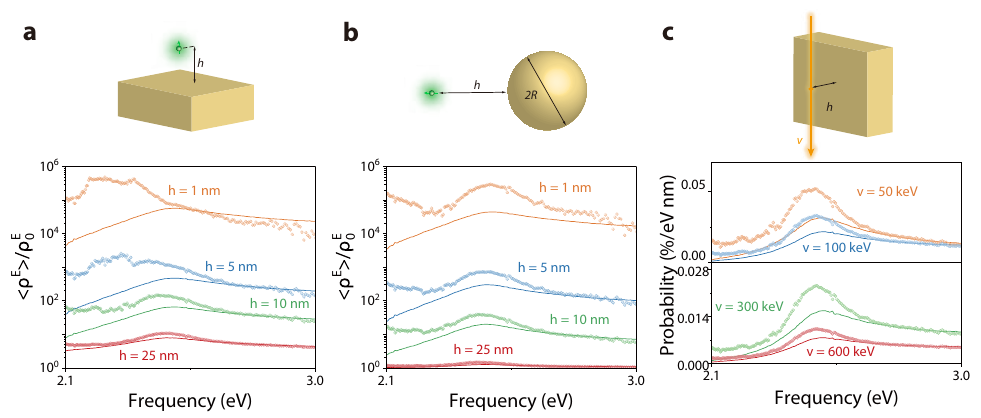}
    \caption{\textbf{Quantum corrections to the local density of states and electron energy loss.}
    \textbf{a.} LDOS near a planar gold surface at different separations $h$.
    \textbf{b.} LDOS at a different separation of $h$ from a gold sphere with a fixed radius $R = \SI{5}{\nm}$.
    \textbf{c.} Electron energy loss of different velocities above at a fixed separation of $h=\SI{5}{\nm}$ above a planar gold surface.
    Classical and nonclassical calculations are denoted by solid lines and dots, respectively.
    We use the measured bulk permittivities and $d$-parameters for the calculations.
    }
    \label{smfig:enhancement}
\end{figure}

\subsection{Free electron energy loss near a planar surface}
\label{smsec:eels}
Assuming the surface at $z=0$ and the velocity of the electron is $\mathbf{v}= v\mathbf{\hat{x}}$ with distance $h$ above the surface, the electron energy loss probability per unit length is 
\begin{equation}\label{smeq:eels}
    \frac{\diff \Gamma\left(\omega\right)}{\diff x}=\frac{2\alpha}{\pi c \beta^2}\int_0^\infty\frac{\diff k_y}{q^2}\,\Re\left\{\e^{\iu 2 k_{z,1}h}\left[ \frac{k_y^2 \beta^2}{k_{z,1}}r_\mathrm{s}-\frac{k_{z,1}}{\eps_1}r_\mathrm{p}  \right] \right\},
\end{equation}
where $\beta= v/c$, $\alpha \approx 1/137$ is the fine-structure constant, and $k_y = \sqrt{q^2-(\omega/v)^2}$. Again, modified reflection coefficients contain the nonclassical contributions described by $d$-parameters~\cite{gonccalves2023interrogating}. 
As Fig.~\ref{smfig:enhancement}c shows, the smaller the velocity of the free electrons, the larger the nonclassical enhancement. This is because electrons of smaller velocity couple to large momenta $q$, where the $\dperp$ contribution scales as $q^2\dperp$.

\subsection{Free electron energy loss and cathodoluminescence near a sphere}
\label{smsec:eels_sphe}

We propose another way to experimentally probe the Bennett mode by using high-resolution electron energy loss spectroscopy. Consider a point electron beam passing near a gold sphere, the corresponding electron energy loss and cathodoluminescence (CL) can be derived based on the Mie theory \cite{de1998relativistic,de1999relativistic,gonccalves2023interrogating}. The EELS and CL probabilities are given by
\begin{align}
    \Gamma_{\mathrm{EELS}}(\omega) =& \frac{\alpha}{\omega\sqrt{\eps_\mathrm{d}}}\sum_{l=1}^{\infty}\sum_{m=-l}^{l}\, K_m^2\left(\frac{\omega \left(R+h\right)}{v\gamma_{\eps_\mathrm{d}}}\right)\times \left[C_{lm}^{\mathrm{p}}\left(\beta_{\eps_\mathrm{d}}\right)\Im\left\{a_{\mathrm{p},l}\right\}+C_{lm}^{\mathrm{s}}\left(\beta_{\eps_\mathrm{d}}\right)\Im\left\{a_{\mathrm{s},l}\right\}\right], \\
    \Gamma_{\mathrm{CL}}(\omega) =& \frac{\alpha}{\omega\sqrt{\eps_\mathrm{d}}}\sum_{l=1}^{\infty}\sum_{m=-l}^{l}\, K_m^2\left(\frac{\omega \left(R+h\right)}{v\gamma_{\eps_\mathrm{d}}}\right)\times \left[C_{lm}^{\mathrm{p}}\left(\beta_{\eps_\mathrm{d}}\right)\left|a_{\mathrm{p},l}\right|^2+C_{lm}^{\mathrm{s}}\left(\beta_{\eps_\mathrm{d}}\right)\left|a_{\mathrm{s},l}\right|^2\right],
\end{align}
where $K_m$ is the modified Bessel function of the second kind, $\beta_{\eps_\mathrm{d}}=\sqrt{\eps_\mathrm{d}}v/c$, and the Lorentz factor $\gamma_{\eps_\mathrm{d}}=\left(1-\beta_{\eps_\mathrm{d}}^2\right)^{-1/2}$. 
$C_{lm}^{\mathrm{p}}$ and $C_{lm}^{\mathrm{s}}$ only relate to the electron beam energy (see Ref.\cite{garcia2010optical} for explicit expressions). 
The quantum correction enters via the Mie coefficients $a_{\mathrm{p},l}$ and $a_{\mathrm{s},l}$ whose explicit forms are Eq.\eqref{smeq:ldos_sphere_mie}. 
Our predictions are shown in Fig. 4b in the main text. The peaks in the quantum corrections exhibit blueshift compared with the classical results. Two peaks emerge in the EELS spectra, with the Bennett mode on the low-frequency side.

\section[Quantum corrections to field emission]{Field emission mediated by quantum surface response}
\label{smsec:fieldemission}

Here we make new predictions on how field emission could be enhanced by quantum surface response.
The principle of field emission is widely used in many applications, especially electron guns in modern electron microscopes. 
Field emission strongly depends on the work function, which is the difference between the energy level infinitely far in the contacted material (usually assumed vacuum) and the Fermi level of electrons in solids.
Here, we show that quantum surface responses driven by external incident light can modify the work function, leading to enhanced or suppressed emission depending on whether $\Re\dperp$ is positive or negative.

The derivation is divided into two parts. We first explain how the $d$-parameter-mediated field discontinuity modifies the work function, which is then inserted into the Fowler--Nordheim framework to calculate the field emission current.

\subsection{Work function changes with the strong electric field}
\label{smsec:workfunction}
In the context of $d$-parameters, nonclassical surface dipoles $\boldsymbol{\pi}$ appears under field excitation via Eq.~\eqref{smeq:dipole_define} which we may re-express as
\begin{equation}
        \boldsymbol{\pi}(\mathbf{r}_{\partial\Omega})\equiv \epsz\dperp\jump{E_\perp}\hat{\mathbf{n}}, \label{smeq:dipole_define_2}
\end{equation}

It is well-established that out-of-plane surface dipoles can form a polarization barrier, thereby changing the work function~\cite{smoluchowski1941anisotropy,fleig2005work,vitali2010portrait,kahn2016fermi}. 
Specifically, the work function change $\Delta \phi = \phi-\phi_0$ (where $\phi_0$ is the work function without external fields) due to a surface dipole density is%
~\cite{forbes1978negative,leung2003relationship,ibach2006physics,avila2020role}
\begin{equation}
    \Delta \phi = -\frac{e}{\epsz} \pi,
\end{equation}
where $\pi$ denotes the dipole density along the out-of-plane direction for simplified notation.

More concretely, let us consider a TM incident Gaussian pulse onto an equivalent infinitesimal planar metallic surface (surface normal direction being $z$). The amplitude of the total field along $z$ is
\begin{align}\label{smeq:out_total_field}
    E^+_\perp=F_0\sin{\theta}~\left(\e^{-\iu k_{z,\mathrm{1}}z}+r_\mathrm{p}\e^{\iu k_{z,\mathrm{1}}z}\right)\e^{\iu \left(qx-\omega t\right)}\e^{-t^2/4\sigma^2}.
\end{align}

Here $F_0$ is the enhanced peak field, $\theta$ is the incident angle, $\e^{-t^2/4\sigma^2}$ represents the Gaussian envelope, $\sigma$ describes the duration of the pulse, $k_{z,\mathrm{1}}$ is the out-of-plane wavevector, and $q$ is the in-plane wavevector.
Recalling Eq.~\eqref{smeq:mesoscopic_bcs_Dperp} and again assuming $\dpar=0$, we have
\begin{equation}\label{smeq:Dperprelation}
    \jump{D_\perp} 
    = D_\perp^+-D_\perp^-
    = \eps^+E_\perp^+-\eps^-E_\perp^-
    = \dpar\nablav_\para\cdot\jump{\mathbf{D}_\para}=0.
\end{equation}

Thus, the relation between the modified work function and the incident field at the $z=0$ surface is given by 
\begin{align}\label{smeq:wf}
    \Delta \phi=-\frac{e}{\epsz}\Re\left[\pi\right] = -\frac{e}{\epsz} \cdot \Re\left[\dperp \epsz \frac{\eps^- - \eps^+ }{\eps^- }E_\perp^+\right]=-eF_0\Re\left[\dperp\frac{\eps^--\eps^+}{\eps^-}\sin{\theta}~\left(1+r_\mathrm{p}\right)~\e^{\iu\left(qx-\omega t\right)}\e^{-\frac{t^2}{4\sigma^2}}\right].
\end{align}
For simplicity, we consider the average change of work function within the entire pulse duration:
\begin{align}\label{smeq:average_wf}
    \langle\Delta \phi\rangle=\Biggl\langle-eF_0\Theta(-F(t))\Re\left[\dperp\frac{\eps^--\eps^+}{\eps^-}\sin{\theta}~\left(1+r_\mathrm{p}\right)~\e^{\iu \left(qx-\omega t\right)}\e^{-\frac{t^2}{4\sigma^2}}\right]\Biggr\rangle.
\end{align}
Here $\Theta[-F(t)]$ is the Heaviside function to ensure that only half-cycles of $F(t)<0$ are considered, during which electrons can be pulled outside the material. 
Therefore, electron spill-out results in a negative surface dipole layer, causing the work function to increase. On the contrary, electron spill-in leads to a positive surface dipole layer and reduction of work function.
The averaged change of the work function is shown in Fig.~\ref{smfig:fieldemission}a.
The resulting Keldysh parameter~\cite{keldysh1965ionization} from Eq.~\ref{smeq:average_wf} is on the order of unity, which enables us to model the electron field emission in the Fowler--Nordheim regime. 

\subsection{The field emission current}

The Fowler--Nordheim model describes the field emission current escaping from the solid surface when a strong field is applied. Here we adopt the Fowler--Nordheim equation in the form of~\cite{fowler1928electron,keathley2019vanishing,jensen2024tutorial} 
\begin{equation}\label{smeq:FNexpression}
    J_{\mathrm{FN}}=\frac{e^3}{16\pi^2\hbar} \frac{[F(t)]^2}{\phi}\Theta[-F(t)]\e^{\frac{4\sqrt{2m\phi^3}}{3\hbar e F(t)}},
\end{equation}
where $F(t)=\Re(E_\mathrm{inc})$ is the incident field. The average current density is~\cite{putnam2017optical}
\begin{align}
    J = f_\mathrm{R}\int^{T_\mathrm{R}/2}_{-T_\mathrm{R}/2}J_{\mathrm{FN}}(t)\mathrm{d} t,
\end{align}
where $T_\mathrm{R}=1/f_\mathrm{R}$ and $f_\mathrm{R}$ is the repetition rate. Figure~\ref{smfig:fieldemission}b shows that the current enhancement or suppression depends strongly on $\Re\dperp$ and incident angles. 
It is found that $\Re\dperp>0$ leads to current suppression and $\Re\dperp<0$ leads to current enhancement.

\begin{figure}[htbp]
    \centering
    \includegraphics[width=0.8\linewidth]{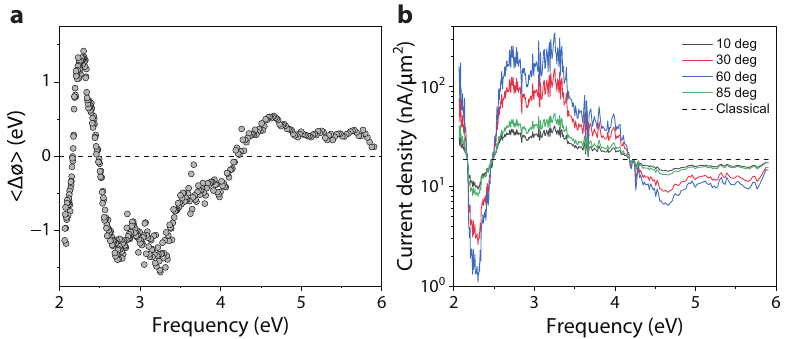}
    \caption{\textbf{Modified work function under quantum surface corrections and the resulting field emission current density.}
    \textbf{a.} Modified work function under pulsed laser pump at different center frequencies. Here the original work function is chosen as $\phi_0=\SI{5.1}{\eV}$, the peak field is \SI{15}{V/nm}, the repetition rate $f_\mathrm{R}=\SI{100}{\MHz}$ and the full-width at half-maximum duration is \SI{10}{\fs} \ which contains enough number of optical cycles such that the phase of the pulse plays a negligible role here. 
    \textbf{b.} Predicted field emission current density under various incident angles. 
    }
    \label{smfig:fieldemission}
\end{figure}

Moreover, the incident angle also has a large impact on the enhancement because of the balance between two counteractive factors.
First, smaller incident angles lead to smaller electric field components (and their discontinuity) along the normal $z$ direction, resulting in weakened modifications (either enhancement or suppression) to field emission. 
Second, toward larger incident angles close to \SI{90}{\degree}, $r_\mathrm{p}\rightarrow-1$, resulting in a vanishing total field.
Jointly the considerations imply an optimal, intermediate incident angle (close to \SI{60}{\degree} for the case shown in Fig.~\ref{smfig:fieldemission}b) that maximizes the nonclassical modifications to field emission.




\bibliographystyle{apsrev4-2}
\bibliography{si}